\documentclass[journal]{IEEEtran}

\ifCLASSINFOpdf
  \usepackage[pdftex]{graphicx}
  \graphicspath{{../pdf/}{../jpeg/}}
  \DeclareGraphicsExtensions{.pdf,.jpeg,.png}
\else
  \usepackage[dvips]{graphicx}
  \graphicspath{{../eps/}}
  \DeclareGraphicsExtensions{.eps}
\fi

\usepackage{amsmath}
\usepackage{color}
\usepackage{amssymb}
\DeclareMathOperator*{\argmin}{argmin}

\usepackage{svg}
\usepackage{enumerate}

\usepackage{algorithm}
\usepackage[noend]{algpseudocode}

\ifCLASSOPTIONcompsoc
\usepackage[caption=false,font=normalsize,labelfont=sf,textfont=sf]{subfig}
\else
\usepackage[caption=false,font=footnotesize]{subfig}
\fi

\usepackage{textcomp}

\begin{document}

%
\title{Experimental Automatic Calibration of a \\Semi-Active Suspension Controller\\via Bayesian Optimization}

\author{Gianluca~Savaia,
        Youngil~Sohn,
        Simone~Formentin,
        Giulio~Panzani,
        Matteo~Corno,
        and~Sergio~M.~Savaresi
\thanks{G. Savaia, S. Formentin, G. Panzani, M. Corno and S. M. Savaresi are with the Department of electronics, information and bioengineering, Politecnico di Milano, Italy. Y. Sohn is with Hyundai Motor Compay, Seoul, Republic of Korea. Corresponding author: {\tt simone.formentin@polimi.it}, Piazza Leonardo da Vinci 32, 20133, Milano, Italy.}
}



\maketitle

\begin{abstract}
The End-of-Line (EoL) calibration of semi-active suspension systems for road vehicles is usually a critical and expensive task, needing a team of vehicle and control experts as well as many hours of professional driving. In this paper, we propose a purely data-based tuning method enabling the automatic calibration of the parameters of a proprietary suspension controller by relying on little experimental time and exploiting Bayesian Optimization tools. A detailed methodology on how to select the most critical degrees of freedom of the algorithm is also provided. The effectiveness of the proposed approach is assessed on a commercial multi-body simulator as well as on a real car. 
\end{abstract}


\IEEEpeerreviewmaketitle

\section{Introduction} \label{sec:intro}
\IEEEPARstart{T}{he} fine tuning of semi-active suspension systems \cite{savaresi2010semi} for road vehicles is known to be a costly and burdensome task, usually needing a close collaboration among professional drivers, control system experts and test engineers, as well as many hours of driving; furthermore, the evaluation of the riding comfort often relies on human sensibility which is hardly quantifiable. For these reasons, in \cite{milliken1995race} the tuning of the damper was defined as a \emph{black art}.

A popular option to make EoL calibration an easier and more efficient task is to rely upon \textit{Hardware-in-the-Loop} (HiL) simulators, where the driver seats in the car cockpit and drives within a virtual environment, while the variations of the road profile are emulated by a robotic arm, see \cite{schuette2005hardware}. Another alternative to real-world tests is the so-called \textit{4-poster test system}, where the car is placed on four moving platforms that emulate the road excitation \cite{chindamo2017reproduction}. Despite their major advantages in terms of costs and repeatability of the tests, such procedures are clearly limited by the lab environment and still require domain experts to perform a suitable calibration.

For the above reasons, in this paper we propose a data-driven calibration method, in which the problem is formulated as a \emph{sequential decision making} task, see \cite{barto1989learning}, allowing the test engineer to perform a fully automatic tuning of the suspension system directly on the car. The input of the algorithm will only be the data streamed by the on-board sensors, while the control parameters will be returned as outputs with no need for human intervention, by minimizing a pre-defined performance-oriented objective function (\emph{e.g.}, addressing comfort or road handling) via \textit{Bayesian Optimization} (BO) \cite{snoek2012practical, klein2016fast}. We will argue that, using the latter tool, only a limited amount of driving time is required, thus making real experimental tests cost-competitive with respect to HiL and other indoor alternatives. This would not be true for other global optimization approaches, \emph{e.g.}, Particle Swarm or Genetic Algorithms, since they usually require a much higher number of experiments, as discussed in \cite{alkhatib2004optimal}. For more details on Bayesian Optimization, see the surveys in \cite{shahriari2015taking, frazier2018tutorial, brochu2010tutorial}. 

In the recent years, similar approaches to the one presented in this contribution, all based on Gaussian Processes \cite{villemonteix2009informational}, have been successfully investigated in the robotic field \cite{marco2016automatic, marco2017virtual}, quadrotors \cite{berkenkamp2016safe}, power plants \cite{abdelrahman2016bayesian} and PID control \cite{lygeros2020cascade}.

To summarize, the contribution of this work consists of a fully automatic calibration protocol of a fixed-structure suspension controller for a road vehicle. The protocol defines the working conditions which the test must follow, the definition of the objective function that best represents the major key performance index and a BO-based sampling strategy, which suggests the calibration parameters which shall be tested to rapidly converge to the optimum. A preliminary version of this work, with a simplified version of this procedure, can be found in \cite{savaia2020bo} where the proof-of-concept was performed in a simulation environment only, employing a simple control logic consisting of only two parameters. In addition, in this contribution, the methodology is expanded to consider a realistic controller and it is validated on an experimental vehicle.

The remainder of this paper is as follows. In Section \ref{sec:problem}, the problem is formally stated for the case of a parametric semi-active suspension controller. The proposed BO-based automatic tuning algorithm is illustrated in detail in Section \ref{sec:methodology}. The experimental setup is described in Section \ref{sec:setup}, while Section \ref{sec:experiments} shows the calibration procedure as well as the obtained results on the considered setup. The paper is ended by some concluding remarks.


\section{Problem Statement} \label{sec:problem}
A commercial semi-active suspension controller is a very complex software which may consist of hundreds of tunable parameters. This magnitude of complexity makes the calibration a burdensome task which requires a significant amount of resources, in particular: 
\begin{itemize}
    \item a calibration engineer who needs to have a good understanding of the software and the effect of control parameters on vehicle dynamics;
    \item a professional driver who needs to evaluate the vehicle dynamics with respect to the proposed calibration.
\end{itemize}

The fine-tuning of a production vehicle may take months also in the case the calibration engineer is accustomed to the control software, since he shall consider many different aspects concerning the calibration of a suspension system; for more insights, see \cite{savaia2020sh}. Furthermore, it is economically expensive to perform experiments due to the high costs of the proving ground, human resources and vehicle maintenance.

\begin{figure}
    \centering
    \subfloat[]{\includegraphics[width=\columnwidth]{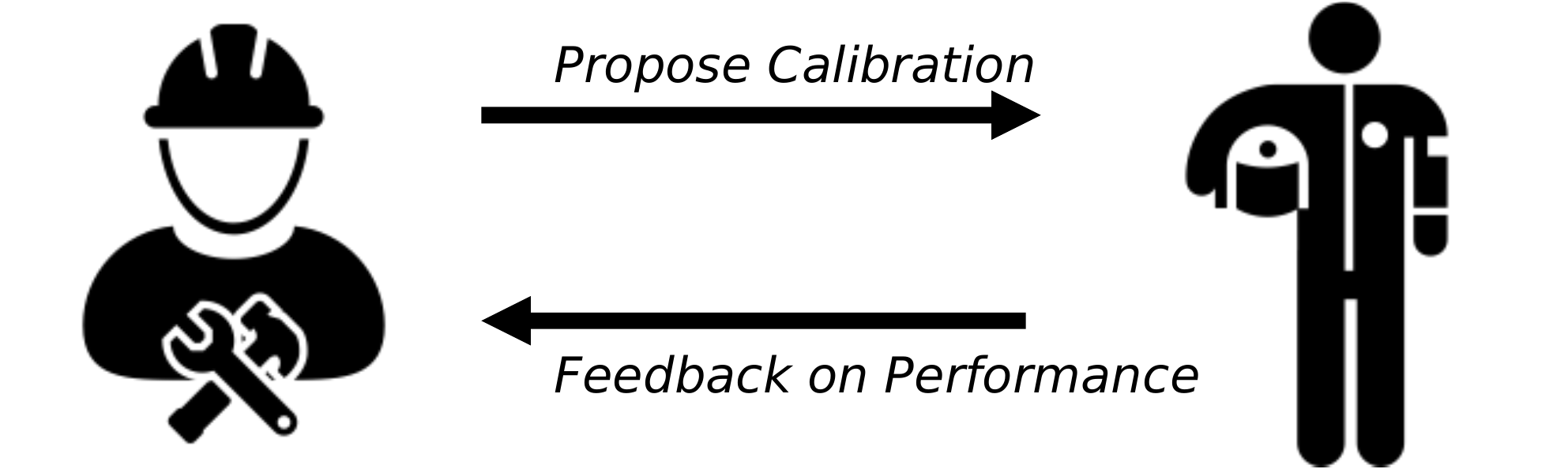} \label{fig:scheme_human}}
    \hfil
    \subfloat[]{\includegraphics[width=\columnwidth]{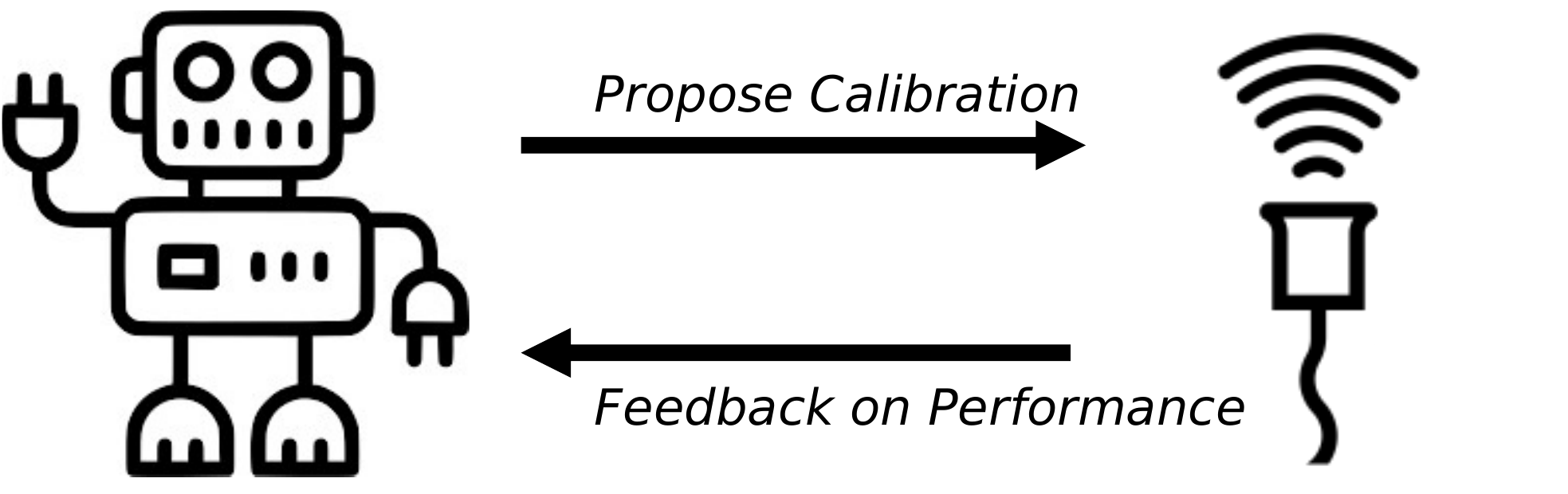} \label{fig:scheme_automatic}}
    \hfil
\caption{The present human-in-the-loop (a) and the proposed automatic (b) calibration paradigms.}
\label{fig:scheme}
\end{figure}

The objective of this research consists in proposing a methodology for the automatic calibration of a suspension control software, shifting from a human-in-the-loop paradigm \figurename \ref{fig:scheme_human} to a fully automatic paradigm \figurename \ref{fig:scheme_automatic}: the calibration engineer is substituted with a RL agent and the driver's feedback by the sensing units equipped on-board.

We argue that the presented methodology can calibrate autonomously the software directly on the vehicle using less resources, time and money than the current \emph{human} paradigm.

\section{The proposed methodology} \label{sec:methodology}

This section introduces the proposed Bayesian Automatic Calibration (BAC) protocol for a parametric suspension control software. The foundations of this methodology, based on Bayesian Optimization, were presented in \cite{savaia2020bo}. In this work, we consider a realistic scenario with a real-world setup. This activity of ``bridging the gap'' between theory and practice required a careful review of the design of experiments as well as the choice of the degrees of freedom involved in the calibration. 

The complexity of dealing with a realistic control software is tackled in two separate phases:
\begin{itemize}
    \item a \emph{simulation phase} aimed at identifying a subset of the most significant parameters;
    \item an \emph{experimental phase} aimed at searching the optimal values of the calibration parameters directly on the real vehicle.
\end{itemize}

The latter point is important because we work with the assumption that the simulation environment is not a perfect digital twin of the real scenario: what is observed in simulation may not be replicable on the real vehicle.

On the other hand, experiments on a real vehicle are expensive (time-wise and money-wise) and shall be kept as few as possible. For this reason, the simulation's aim is important: the less parameters needing calibration, the less experiments are needed to reach the optimum calibration on the real vehicle. 


\subsection{Experimental Protocol}
The experiments shall be performed on a single test road representative of the dynamics of interest. This is a critical point, since the result of the optimization strongly depends on the road excitation; in other words, the optimal configuration for a driving cycle on a highway and off-road are expected to differ significantly. Hence, the choice of the road profile has an important impact on the outcome of the optimization and shall be carefully evaluated in the design phase.

Each experiment must be performed over the same road segment, following the same velocity profile, in order to ensure the same frequency excitation.

\subsection{Key Performance Index} \label{sec:kpi}
The definition of the performance index is an important ingredient of the methodology, since the output calibration will be a direct reflection of this definition.

A standard key performance index of comfort due to mechanical vibrations can be found in the document ISO 2631-1 \cite{ISO2631}, where the vehicle dynamics is evaluated in six degrees of freedom (translational and rotational) with respect to the human perception of discomfort (\emph{i.e.}, motion sickness) based on heuristics.

In the industry, this standard is seldom employed in the performance evaluation of a suspension system. Instead, most of car manufacturers rely on subjective evaluation of the vehicle dynamics; indeed, each vehicle has a distinguished signature characterizing its dynamics, though some common features are always common (\emph{e.g.}, vertical stability). The test driver must be highly skilled and must have a long experience on the particular signature the car manufacturer wants to imprint to its vehicles.

In the proposed methodology, the driver is replaced by a sensing unit; the transfer of the driver's evaluation into a mathematical formulation is an exciting research field which is out of the scope of this article.

In this work, a simplified variant of the ISO Index (based on the root mean square of measurements) is employed, which focuses on the two main variables of interest: i) minimizing the vibrations perceived on the chassis by the passengers and ii) minimizing the pitch movement of the vehicle. This index can be formulated as 

\begin{equation}
    J = \sqrt{\frac{1}{T} \int_0^T A_z^2 dt} + \sqrt{\frac{1}{T} \int_0^T \dot{\theta}^2 dt}
    \label{eq:kpi}
\end{equation}
where $A_z$ and $\dot\theta$ are the vertical acceleration and pitch rate of the vehicle respectively of an experiment of duration $T$; both quantities can be easily measured directly on the vehicle by means of an Inertial Measurement Unit (IMU) installed on the vehicle chassis. We believe that the optimization of this index is a fair proxy of a comfortable vehicle ride.

We would like to remark that the methodology holds for any choice of the performance index, which is subjective and application dependent. In an industrial application, it could be useful to include more measurements, such as wheel accelerations and roll dynamics, to capture a broader understanding of the vehicle dynamics.

\subsection{Bayesian Optimization}
Bayesian Optimization is an effective technique to solve optimization problems characterized by expensive experiments. It consists of two main ingredients: a \emph{surrogate function} to estimate the objective function from data and an \emph{acquisition function} to sample the next observation, see \cite{bemporad2020global}.

The \emph{surrogate function} shall be ideally model-free or non-parametric, unless one desires to constrain the objective function to a certain class (\emph{e.g.}, quadratic). In Bayesian Optimization (BO), the objective function is assumed to be drawn from a Gaussian Process (GP), where each observation is a random variable normally distributed and jointly Gaussian with one another. In particular, each pair of observations is bonded by a covariance function encoding the belief that closer points shall have higher correlation than those far apart; typical choices are the Gaussian and M\`{a}tern kernels.

Hence, given a set of observations, it is possible to fit a GP whose posterior mean is the expected estimation of the objective function, and its variance represents a confidence interval for the estimation. This means that, for any given set of control parameters, it is possible to compute the estimated performance index and a goodness indicator of this estimate.

Given this posterior model, there are several options for the \emph{acquisition function} which aims at sampling the next observation where there is a higher probability to find a minimum, according to the posterior mean and variance. The most popular algorithms are \emph{Expected Improvement} and \emph{Upper Confidence Bound} which can be found declined in different variations to tackle the so-called exploration-exploitation trade-off.

\subsection{Simulation Phase}

The main purpose of this \emph{phase} consists in identifying the most significant (tunable) parameters of the software which affect the vehicle dynamics.

In this work, a fully black-box approach is followed where the semantic meaning of the parameters and the controller architecture is unknown to the calibration system; the only knowledge is the number of tunable parameters $P$ indicated by $\theta = \left[\theta^1, ..., \theta^P\right] \in \mathbb{R}^P$. In the case where additional knowledge exists, the output of the parameter reduction discussed in this section could be supervised by a human.

The classification of the most significant parameters follow a novel approach based on Gaussian Process Regression. Firstly, a Bayesian Optimization framework is setup with $N$ experiments, following the protocol previously discussed; $M$ independent optimizations (consisting of $N$ experiments) are performed (for a total of $N \cdot M$ simulations); notice that this is different from performing only one optimization with $N \cdot M$ experiments because in the considered case the exploration is less biased by the observations, since the optimization tends to explore a neighborhood of the estimated global optimum). The outputs of these optimizations are:
\begin{itemize}
    \item $J \in \mathbb{R}^M$, the minimum reached at the end of each optimization;
    \item $\theta \in \mathbb{R}^{M \times P}$, the optimal calibration parameters reached at the end of each optimization; $\theta_i$, where $i\in [0, M-1]$, indicates the optimal parameters for the $i$th optimization;
    \item $X_1, ..., X_M$, the Gaussian Processes estimated at the end of each optimization based on the sampled observations.
\end{itemize} 

The number of experiments $N$ shall be high enough to guarantee the convergence of the optimization to a minimum. If the optimization problem allows only one global minimum, we could expect that $\forall i,k \in [0, M-1]$ holds
\begin{align*}
    J_i &= J_k  \\
    \theta_i &= \theta_k;
\end{align*}
more generally though, one can expect the optimization problem to admit more local minima with a comparable order of magnitude, namely
\begin{equation*}
    |J_i-J_k| < \varepsilon, \quad \forall i,k 
\end{equation*}
each of which corresponds to a different parametrization $\theta$. 

Let us consider the average expectation of the Gaussian Processes defined as
\begin{equation}
J_\mu(\theta) = \frac{1}{M} \sum_{i=0}^M \mathbb{E}(X_i)    
\end{equation}
which is a surrogate function representing an approximation of the key performance index in \eqref{eq:kpi}. Then, one of the $M$ optimal parameters vector is drawn so that
\begin{equation*}
    \bar\theta = \argmin_{i\in[0,M-1]} J_i(\theta)
\end{equation*}
and a sensitivity analysis is performed starting from $\bar\theta$ by letting each parameter vary, one at the time, along its range of definition. This results in $P$ scalar functions $s^i$ defined as
\begin{equation}
    s^i(\theta^i) = J_\mu(\theta^i) = J_\mu([\bar\theta^1, ..., \theta^i, ..., \bar\theta^P])
\end{equation}
where $\theta^i$ is the $i$th element of vector $\theta \in \mathbb{R}^P$ let free to vary, and all the other $k$ elements, so that $k\neq i$, are kept constant and equal to $\bar\theta$.

The functions $s^i$ are employed as a proxy sensitivity index of parameter $\theta^i$ on the performance index $J$: if the maximum excursion of $s^i$ is negligible with respect to the optimum of $J$, then the optimization problem is not very sensitive to $\theta^i$. This idea can be formulated in the following heuristics
\begin{equation}
    \frac{\max(s^i)-J_\mu(\bar\theta)}{J_\mu(\bar\theta)} > T
    \label{eq:heuristics}
\end{equation}
where $T$ is a threshold indicating the tolerated performance loss (with respect to the optimum) to consider one parameter as negligible. Those parameters violating Equation \eqref{eq:heuristics} are considered as non-significant and discarded from the optimization, effectively reducing the number of parameters.

A further parameter reduction can be achieved by joining together those parameters which share the same minimum: if $\bar\theta^a$ and $\bar\theta^b$ are similar, then they can be seen as one.  

Summarizing, the parameter reduction pipeline proposed in this section is made up of two steps:
\begin{itemize}
    \item \emph{significance}, discarding parameters which do not affect  the performance index significantly
    \begin{equation*}
    \hat\theta = \Bigg\{ \theta^i : \frac{\max(s^i)-J_\mu(\bar\theta)}{J_\mu(\bar\theta)} > T\Bigg\},
\end{equation*}

\item \emph{similarity}, grouping those parameters which have a similar optimum value
\begin{equation*}
    \tilde\theta = \Bigg\{ \hat\theta^i: \forall k > i \implies  |\hat\theta^i-\hat\theta^k| < \varepsilon \Bigg\},
\end{equation*}
\end{itemize}
where $\tilde\theta$ is the final subset of \emph{meta-}parameters which shall be calibrated directly on the vehicle (the term \emph{meta}-parameter is used to indicate that $\tilde\theta$ may refer to more than one parameter $\theta^i$ after the \emph{similarity} step).

\subsection{Experimental Phase}
In this \emph{phase}, the objective consists in searching the optimum calibration for the subset of \emph{meta-}parameters identified in the \emph{simulation phase} by performing experiments on the real vehicle. 

A pseudocode summarizing this procedure, presented in \cite{savaia2020bo}, is shown in Algorithm \ref{alg:bo}. The initialization is performed by picking at random (or any \emph{best guess} policy) a set of control parameters $\theta$, which are evaluated by performing an experiment and then computing the performance index defined in \eqref{eq:kpi}. The objective function is then estimated by fitting a Gaussian Process (GP). Eventually, the next set of control parameters are chosen by maximizing the {expected improvement} (EI), which identifies the region in the parameters space, where it is most likely to find the minimum according to such an acquisition function; the parameters are then evaluated by performing another experiment and the performance index is computed and updated accordingly. 

This class of optimization problems does not rely upon a stopping condition since they are based on the assumption that the evaluation of an experiment is expensive and, thus, there shall be a budget cap to the maximum number of experiments to be performed. Hence, the procedure in Algorithm \ref{alg:bo} is repeated until the allowed maximum number of iterations is reached. 

\begin{algorithm}[H]
\caption{Bayesian Automatic Calibration (BAC)} \label{alg:bo}
    \begin{algorithmic}[]
    \State $\theta^{obs} \gets random$
    \State Perform Experiment
    \State Cut \& Align Signals
    \State Evaluate $J^{obs}$ \\
    
    \While{$i < Max Iterations$}
    \State $\mu, \sigma \gets fitGP(\theta^{obs}, J^{obs})$\\
    
    \State $\theta^{next} \gets argmax \; EI(\mu, \sigma)$
    \State Perform Experiment
    \State Cut \& Align Signals
    \State Evaluate $J^{next}$ \\
    
    \State Update $(\theta^{obs}, J^{obs})$ \\
    \EndWhile\label{euclidendwhile}
    \State $\theta^{opt} \gets argmin(\mu)$
    \end{algorithmic}
\end{algorithm}


\section{Setup} \label{sec:setup}
The methodology presented in Section \ref{sec:methodology} consists of two major phases:
\begin{enumerate}[i]
    \item a \emph{simulation phase}, where the subset of most significant parameters is identified;
    \item an \emph{experimental phase}, where this subset is calibrated on the real vehicle.
\end{enumerate}

In this section, a detailed description of the simulation environment and experimental setup is given. The model employed in simulation does not have to be a perfect model of the real vehicle because the proposed methodology is robust against model uncertainties; if the model were perfect, we would not need the \emph{experimental phase}. Hence, the distortions in the simulation environment with respect to reality (\emph{e.g.}, vehicle model and road profile) do not compromise the methodology presented in Section \ref{sec:methodology}.

\subsection{Simulation Environment} \label{sec:setup_sim}
The simulation of the proposed framework is performed on a full-fledged commercial vehicle simulator which can model complex multi-body dynamics.

\begin{figure}
	\centering
	\includegraphics[width=\columnwidth]{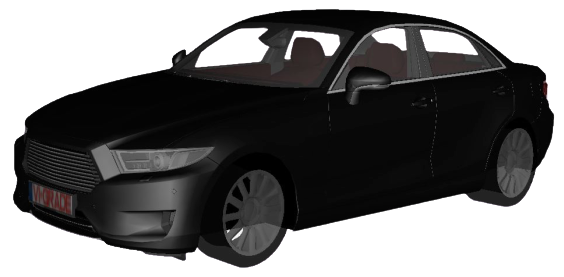}
	\caption{The luxury sedan vehicle employed in simulation.}
	\label{fig:sim_carsim}
\end{figure}

The target vehicle is the luxury sedan illustrated in \figurename \ref{fig:sim_carsim}, where the main parameters of interest are: 
\begin{itemize}
    \item sprung mass: $1370 \, Kg$
    \item unsprung mass: $160 \, Kg$
    \item spring stiffness: $153 \, N/mm$
\end{itemize}

The vehicle is modeled as a MIMO system whose inputs are the damping forces at the four corners and the outputs consists of all the signals characterizing the attitude and status of the chassis, wheels and suspensions.

The control system is implemented in a Matlab/Simulink environment whose outputs consist of the four damping references $C_{ref}$ at the four corners. The actual implementation of the controller is discussed in Section \ref{sec:controllers}.

\begin{figure}
	\centering
	\includegraphics[width=\columnwidth]{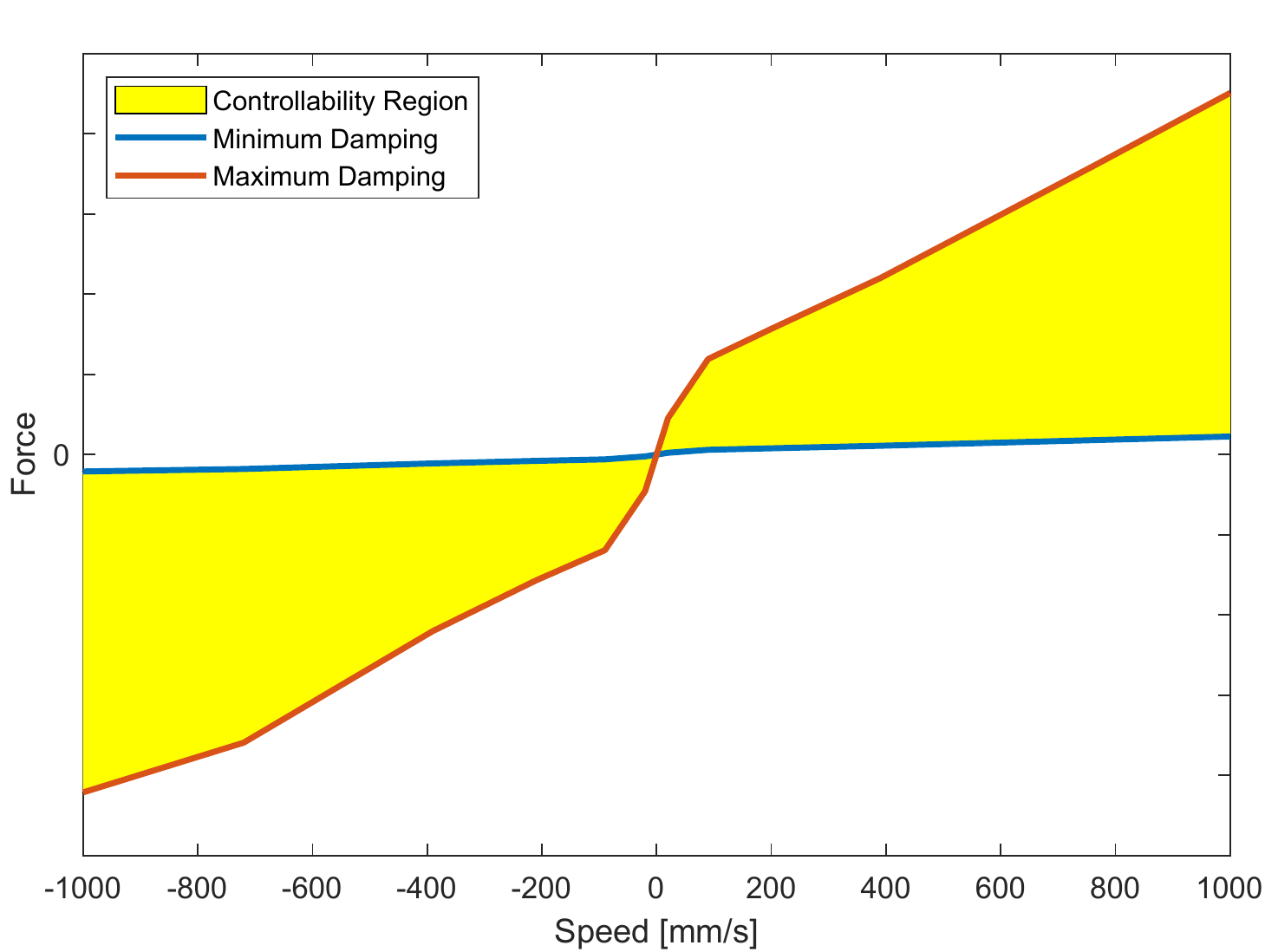}
	\caption{Typical controllability region of a semi-active damper.}
	\label{fig:sim_damping_curves}
\end{figure}

The reference signal $C_{ref}$ is transformed into a damping force $F_d$ by means of the Force - Speed characteristics of the damper, shown in \figurename \ref{fig:sim_damping_curves}; this region defines the boundaries of the minimum and maximum damping force which the device can exert as a function of the suspension's elongation rate $\Delta \dot{z}$. The damping force is therefore given by 
\begin{equation}
    F_d = C_{ref} \Phi_{max}(\Delta \dot{z}) + (1-C_{ref}) \Phi_{min}(\Delta \dot{z}),
    \label{eq:force}
\end{equation}
where $\Phi$ is a non-decreasing function  which shapes the damping characteristics (intrinsic to the physical hardware) and $C_{ref}$ is the control signal which denotes the damping coefficient defined in the range $[0, 1]$ (interpolating between minimum and maximum). For an ideal damper, $\Phi$ should be linear, however for an actual physical device, the damping characteristics looks more similar to the function shown in \figurename \ref{fig:sim_damping_curves}.

These curves do not describe the real damping characteristics of the actual vehicle, but they are a good approximation of what one could expect from such a piece of hardware. As remarked in the previous sections, it is not fundamental that the simulation environment is a perfect model of the real vehicle.

\subsection{Testing} \label{sec:setup_exp}
\begin{figure}
	\centering
	\includegraphics[width=\columnwidth]{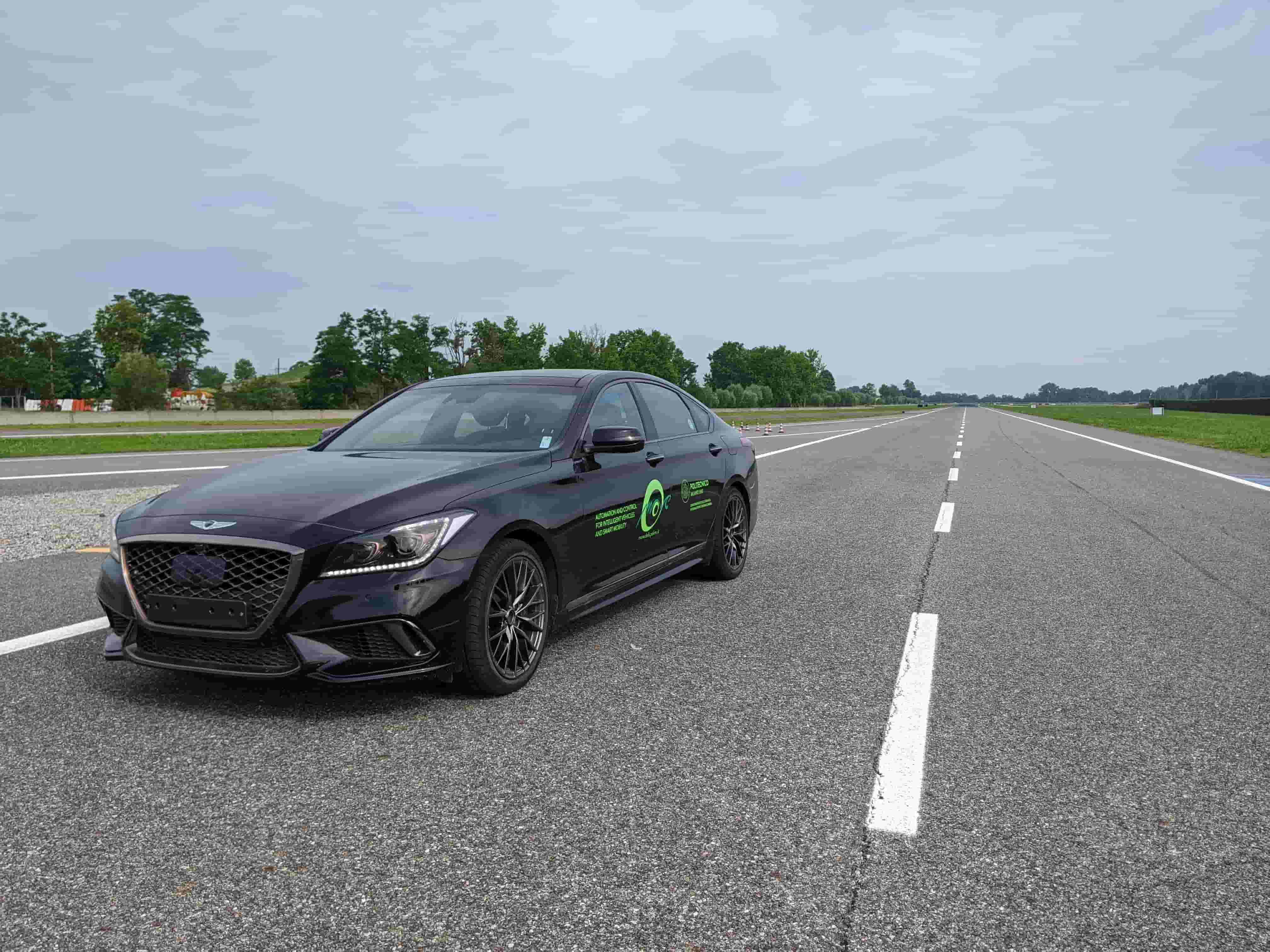}
	\caption{The vehicle employed in the experimental session.}
	\label{fig:exp_car_track_1}
\end{figure}

The experiments are performed on the production vehicle in \figurename \ref{fig:exp_car_track_1} which is equipped with semi-active suspensions.

The control software runs on a prototypal Electronic Control Unit (ECU) which measures three vertical accelerometers on the chassis and two vertical accelerometers on the wheels, and drives the target current to the dampers. The ECU implements both controllers discussed in Section \ref{sec:controllers} which can be selected via a button on the dashboard of the vehicle.

Additionally to the sensors employed for feedback control, an Inertial Measurement Unit (IMU) is installed to the rails of the passenger's seat which measures the vibrations perceived on the chassis and the rotational velocities in order to evaluate the performance index \eqref{eq:kpi}.

The control parameters of the software implemented in the ECU can be tuned real-time via a laptop connected to the vehicle CAN network. This laptop runs Algorithm \ref{alg:bo} described in Section \ref{sec:methodology} implementing the RL agent which sends the suggested calibration to the ECU via a CAN message. This operation is monitored by means of an acknowledgment protocol which ensures the parameters have been correctly delivered to the ECU.

The vehicle is also equipped with a Cruise Control system which is very useful to guarantee a good repeatability of the experiments performed at constant speed.

\begin{figure}
	\centering
	\includegraphics[width=\columnwidth]{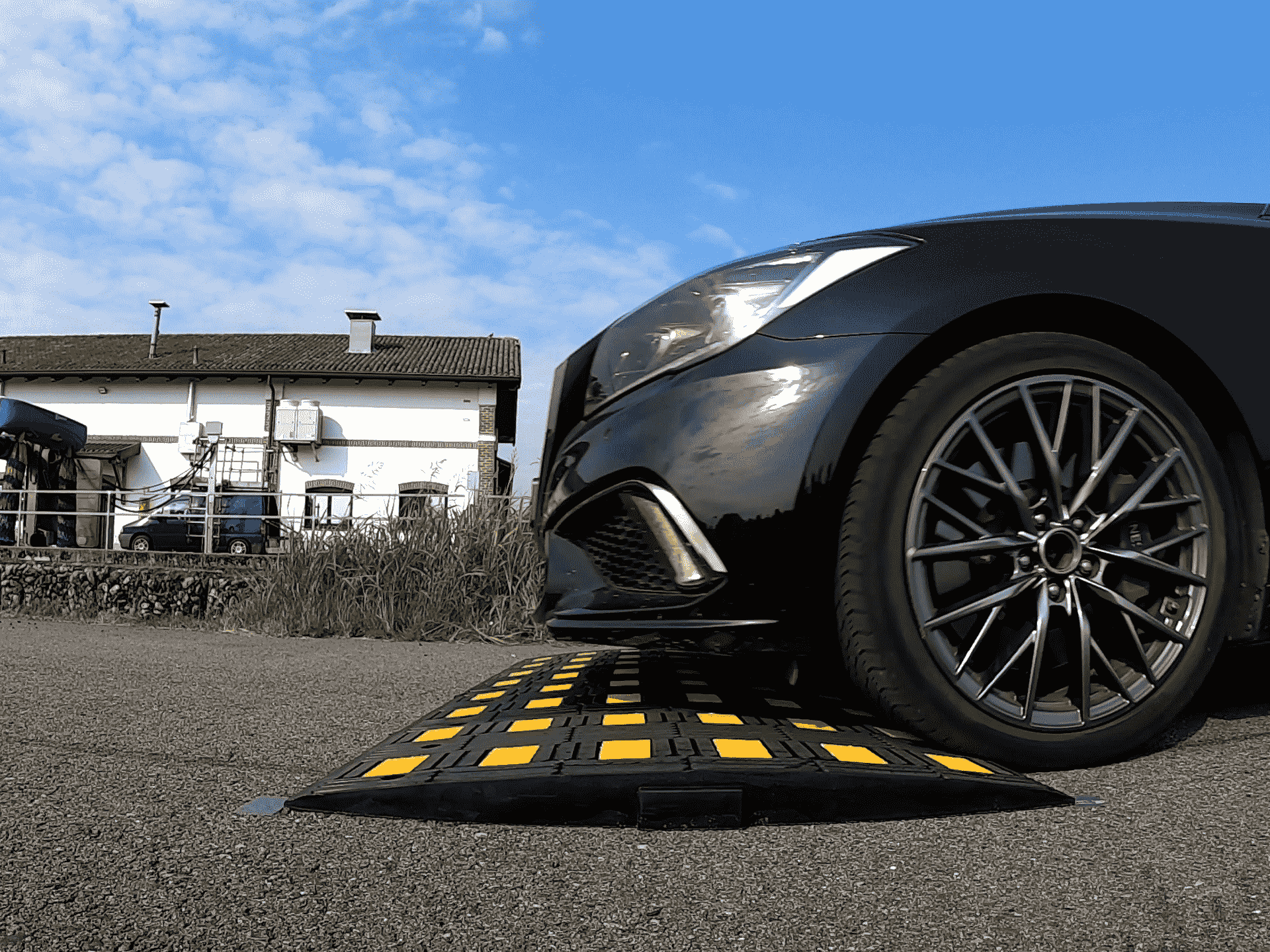}
	\caption{Speed bump.}
	\label{fig:exp_bump_track}
\end{figure}

\begin{figure}
    \centering
    \subfloat[]{\includegraphics[width=\columnwidth]{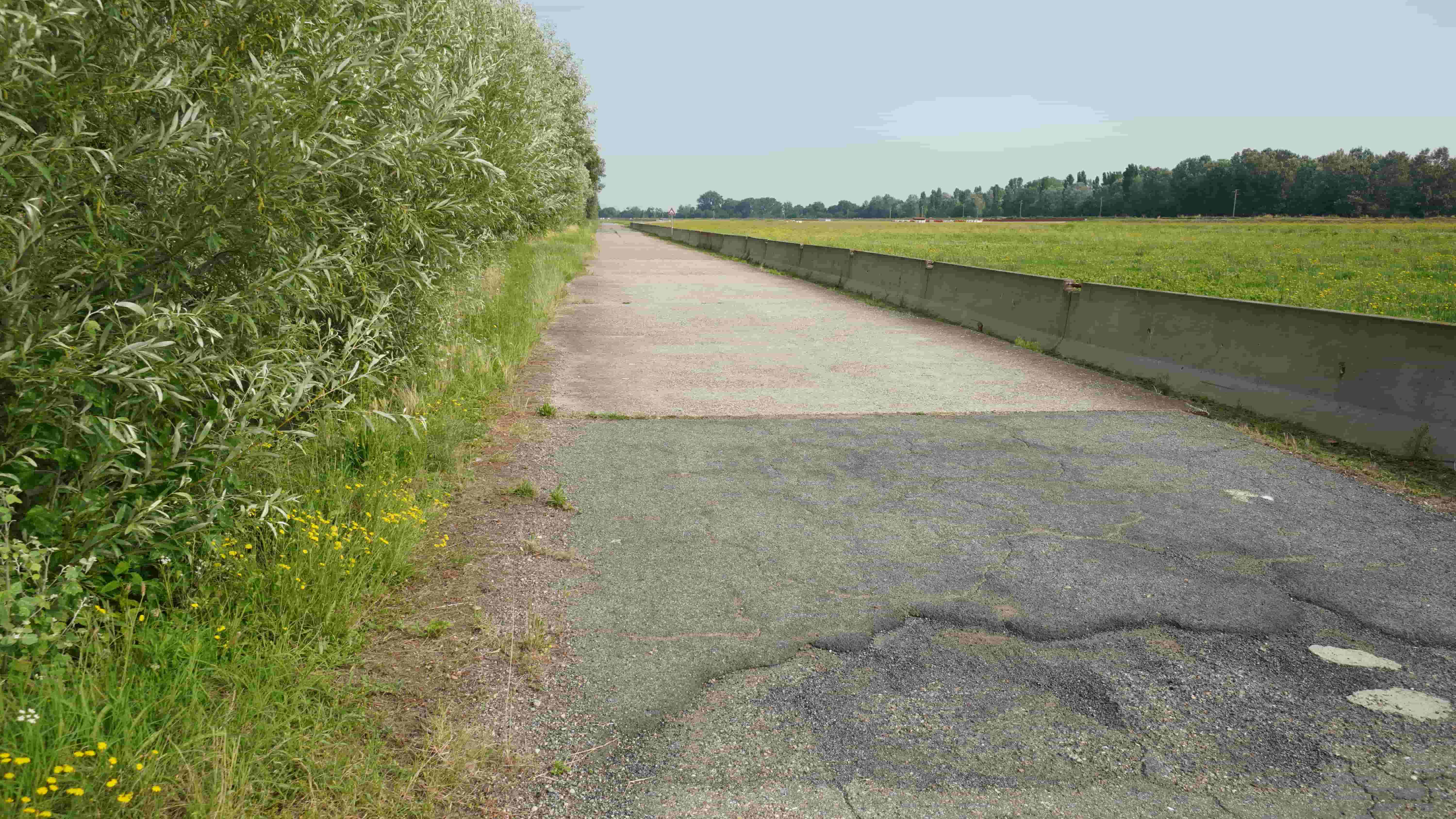} \label{fig:exp_track_1}}
    \hfil
    \subfloat[]{\includegraphics[width=\columnwidth]{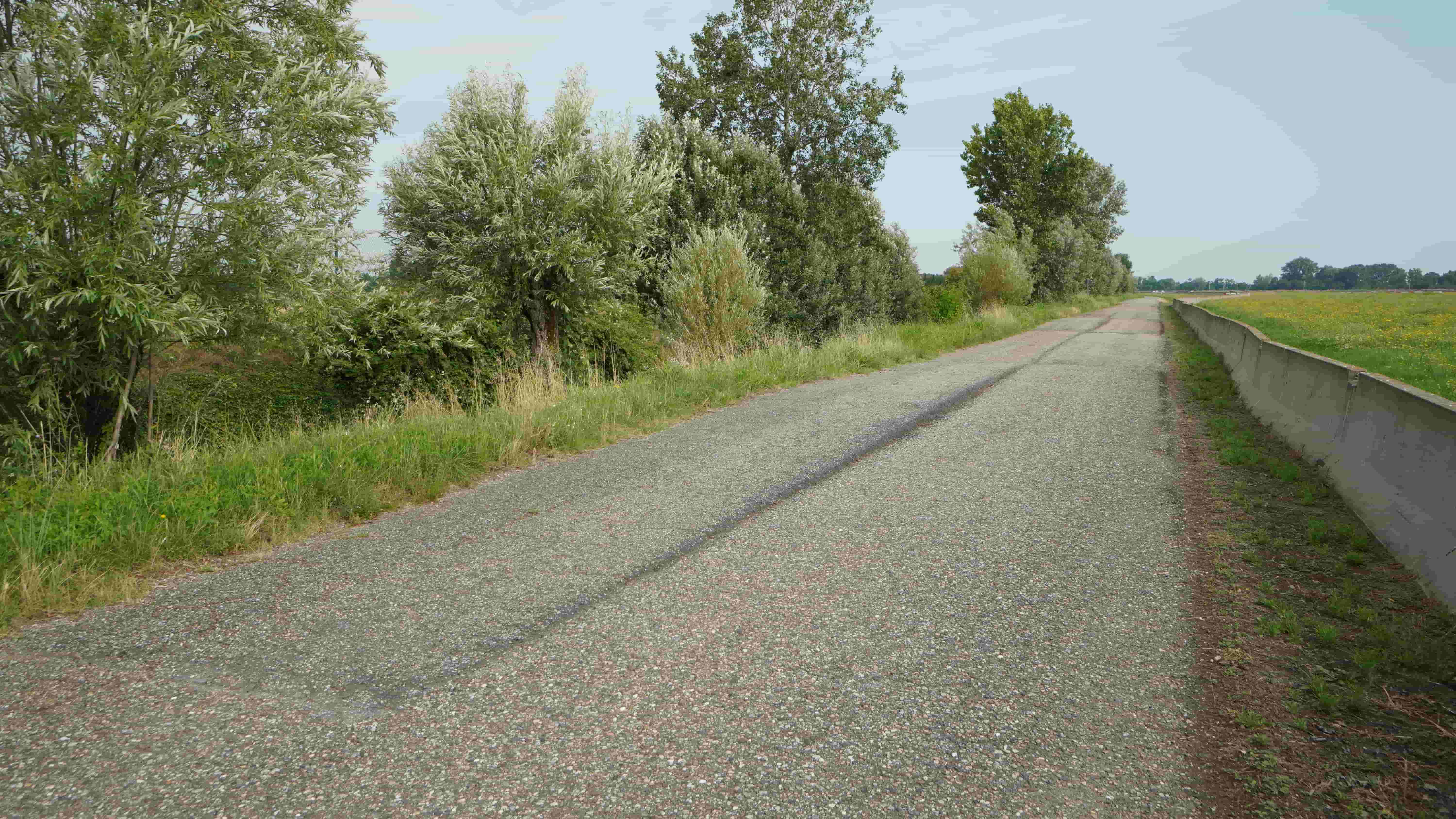} \label{fig:exp_track_2}}
    \hfil
    \subfloat[]{\includegraphics[width=\columnwidth]{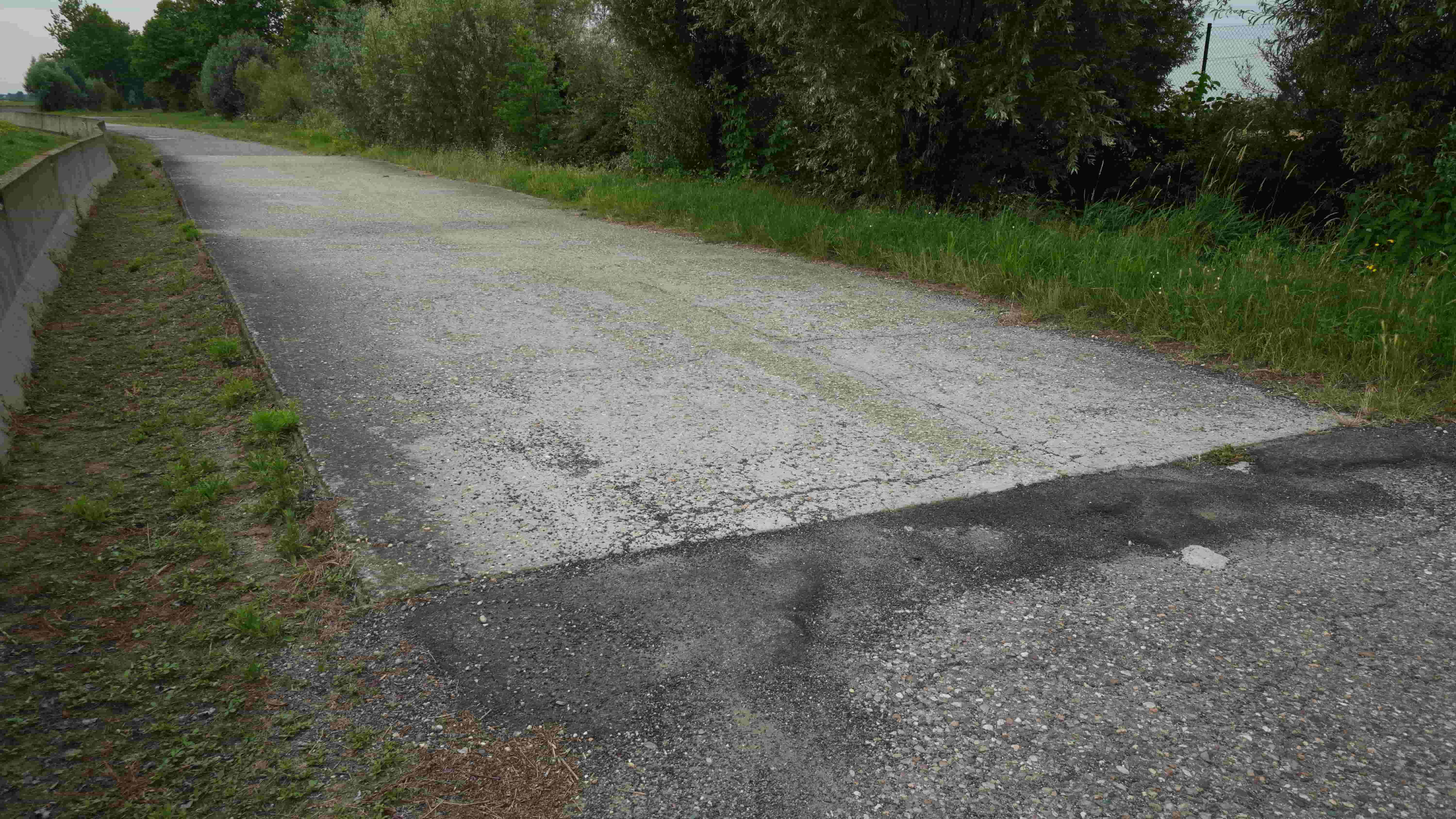} \label{fig:exp_track_3}}
    \hfil
\caption{Test road consisting of rough asphalt (a), asymmetric waves (b) and symmetric waves (c).}
\label{fig:exp_track}
\end{figure}

The experimental validation of the proposed methodology is performed on two different scenarios inside a proving ground:
\begin{itemize}
    \item speed bump, \figurename \ref{fig:exp_bump_track}
    \item test road $1.5km$ long, including low-frequency valleys, high-frequency excitation and asymmetric profiles,  \figurename \ref{fig:exp_track}.
\end{itemize}
These scenarios have been chosen to validate the methodology both on a single event and on a longer stretch of road.


\subsection{Controllers} \label{sec:controllers}
In this work, we present the results of this methodology on two different control software. The first controller in Section \ref{sec:controllers_constant} represents a simplified scenario which is useful to experimentally demonstrate the efficacy of the methodology; the other controller in Section \ref{sec:controllers_semiactive} is an actual industrial software which is implemented on a production vehicle. 

The automatic calibration is performed experimentally on both controllers and discussed separately in the next section.

\subsubsection{Controller I - Constant Damping} \label{sec:controllers_constant}
In this control logic, the dampers are driven with a constant target damping. Therefore, the software consists of two parameters only: (constant) damping on the front and (constant) damping on the rear. The damping is defined in the $(0,1)$ range which corresponds to minimum and maximum damping respectively.

This simple architecture with only two parameters still admits the graphical visualization of the objective function $J$ in a tri-dimensional space; this is very helpful to understand the methodology at work. Furthermore, this small set of parameters gives the possibility to perform an exhaustive search of the optimum calibration; this is important to experimentally prove that the proposed Algorithm \ref{alg:bo} can reach the global optimum.

Despite the simplicity, the optimal calibration of this controller is not trivial since the distribution of the damping between the front and rear axles depends on the geometry of the vehicle and its mass distribution.

\subsubsection{Controller II - Semi-Active Damping} \label{sec:controllers_semiactive}
This control logic exploits the semi-active capabilities of the dampers; it modulates the damping in real-time with respect to the current status of the vehicle which is measured by means of a set of sensors.

For confidentiality reasons, the detailed description of the control logic cannot be disclosed in this article. However, the methodology discussed in Section \ref{sec:methodology} does not require any knowledge on the working principles of the controller following a pure black-box approach. In a real application, it is often the case that either the control logic or (at least) the semantic meaning of the parameters is known; this knowledge can be exploited in the self-calibration algorithm to, \emph{e.g.},  reduce \emph{a priori} the number of significant parameters, reduce the parameters' domain or apply constraints to the optimization.

The only knowledge which is exploited in this work is 
\begin{itemize}
	\item the number of parameters, $N=20$;
	\item each parameter's domain, minimum and maximum value, $0$ and $1000$ respectively.
\end{itemize}

In particular, the semantic meaning is not known which is equivalent to a calibration engineer who has seen the software for the first time and cannot read the parameters' label. We argue that performing the calibration under these conditions would be a challenging task, even for the most experienced engineer.


\section{Experimental results} \label{sec:experiments}

The Constant Damping controller, described in Section \ref{sec:controllers_constant}, serves as a tutorial on the methodology proposed in this article, admitting a graphical interpretation of the results in a three-dimensional space.

In the following, the results on two different scenarios are presented.

\subsection{Speed Bump for Controller I}

The Speed Bump in \figurename \ref{fig:exp_bump_track} is a particularly challenging event for the suspension system whose aim is to absorb the impact, which causes high vertical accelerations, and to dampen the oscillations which follow. There exists a clear trade-off between a low-damping configuration, which performs well in absorbing the impact but poorly for the oscillations damping, and a high-damping configuration, where the opposite holds.

\begin{figure}
	\centering
	\includegraphics[width=\columnwidth]{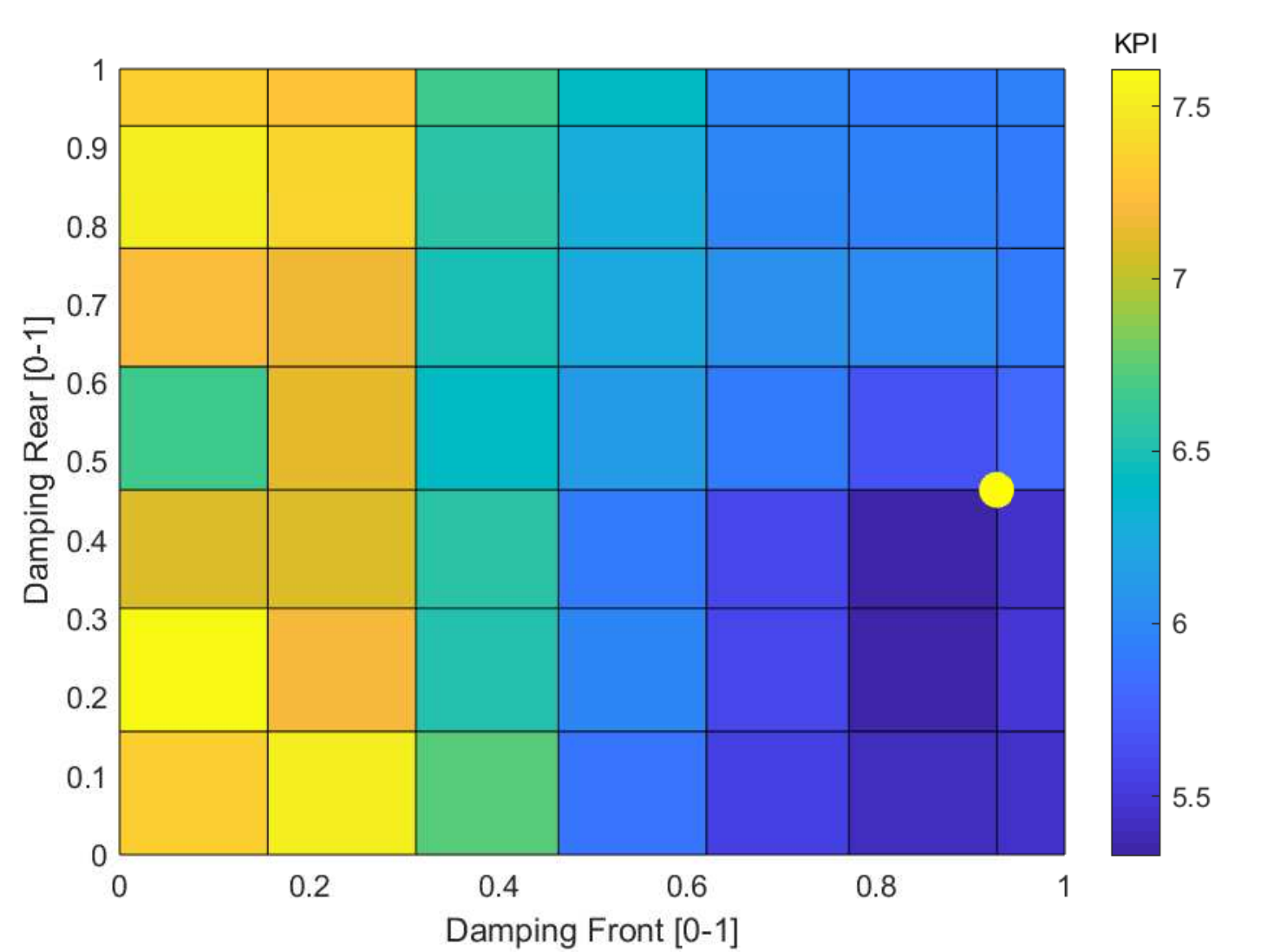}
	\caption{Performance index for \emph{Controller I} on the \emph{Speed Bump} computed via grid search (displayed as contour plot).}
	\label{fig:exp_bump_gridsearch_top}
\end{figure}

The short duration of this test is very convenient from an experimental perspective since it gives the possibility to perform an exhaustive search of the optimum calibration. The parameters space, made up of only two parameters corresponding to the constant front and rear damping, is gridded on a 8-by-8 matrix resulting in a total of 64 experiments. In post-processing, each experiment has been cut and aligned in order to compare the same $2.5s$ time window, and the key performance index in \eqref{eq:kpi} has been computed. The result is the surface representing the index as function of the two calibration parameters as shown in \figurename \ref{fig:exp_bump_gridsearch_top} where, for a better graphical yield, the surface is represented by its projection on the parameters space by exploiting a color code to identify the amplitude of the index; the yellow dot indicates the optimum calibration in the parameters space: the optimum compromise between impact absorption and oscillation damping is obtained by the configuration with a high damping on the front and mid-way damping on the rear.

The result of this grid search analysis is very important because it allows us to have a complete understanding of the performance index which we aim to optimize; this result will be used as ground truth of the output calibration proposed by our data-driven methodology.

\begin{figure}
	\centering
	\includegraphics[width=\columnwidth]{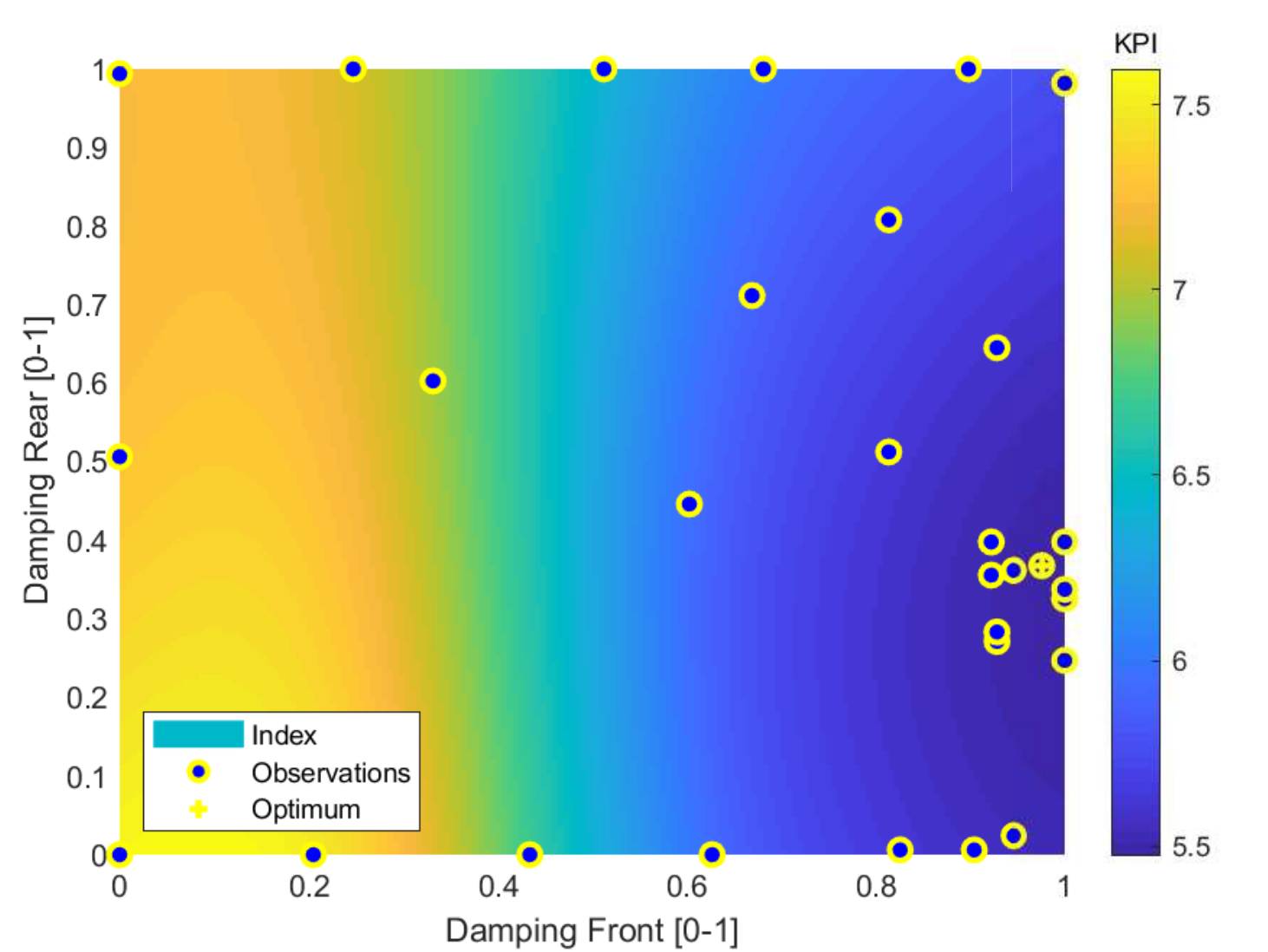}
	\caption{Estimation of the performance index for \emph{Controller I} on the \emph{Speed Bump} via a Gaussian Process.}
	\label{fig:exp_bump_bo_long}
\end{figure}

In this scenario, employing a controller with only two parameters, the \emph{simulation phase} aimed at reducing the number of parameters, as discussed in Section \ref{sec:methodology}, is not needed. Therefore, the parameters can be directly calibrated on the real vehicle by iteratively driving over the Speed Bump at the same constant speed of $30 km/h$. After each experiment, the algorithm automatically cuts and aligns the signals of interest, computes the performance index, estimates the Gaussian Process from the observed data and eventually suggests a new calibration to be tested on the next iteration. The final result is shown in \figurename \ref{fig:exp_bump_bo_long} where the index follows the same color code presented for the exhaustive search, the blue-and-yellow circles indicates the proposed calibrations and the green cross identifies the optimum calibration observed throughout the experiments.

\begin{figure}
	\centering
	\includegraphics[width=\columnwidth]{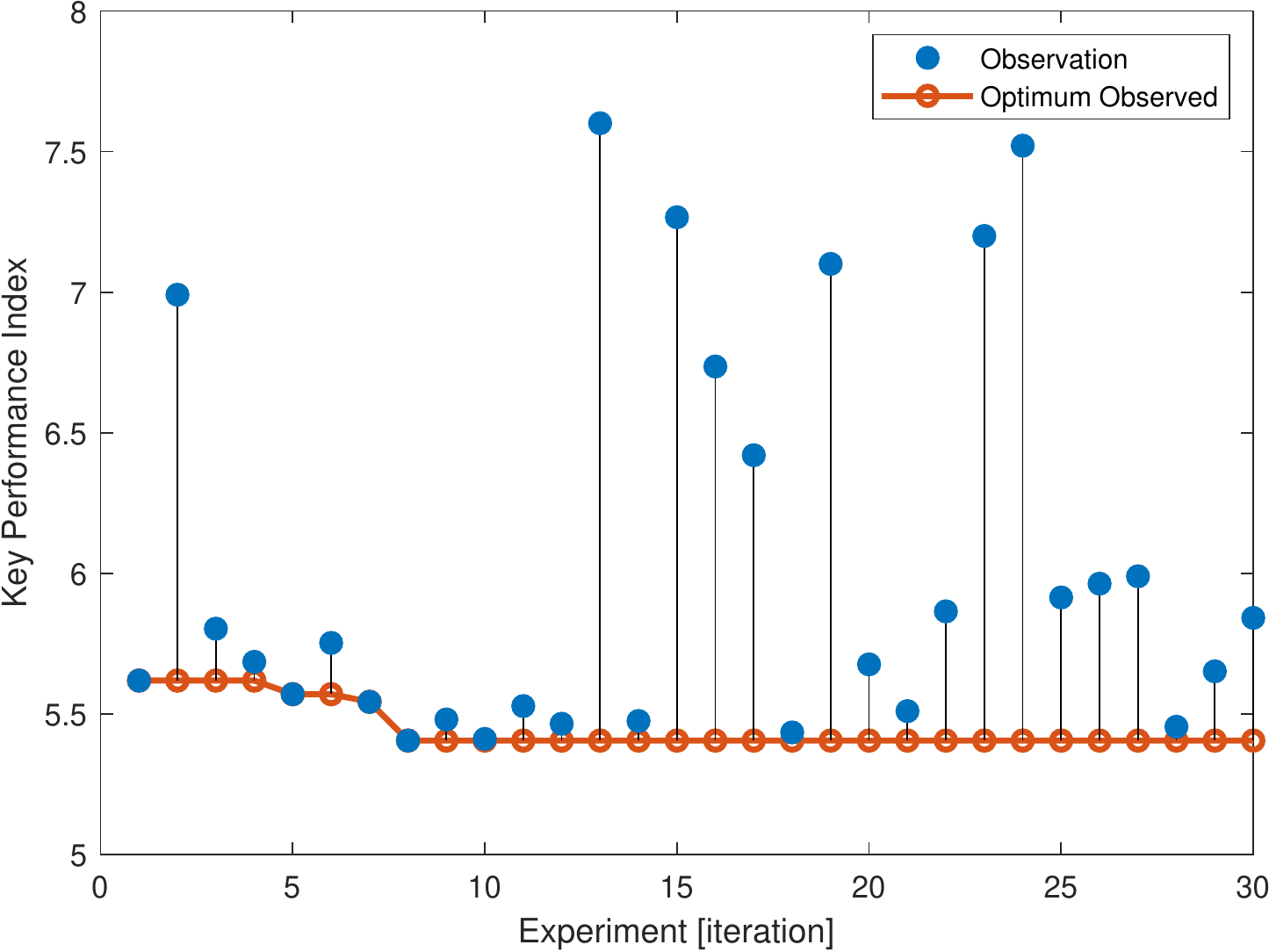}
	\caption{History of the performance index for each experiment for \emph{Controller I} on the \emph{Speed Bump}.}
	\label{fig:exp_bump_bo_long_history}
\end{figure}

The shape of the surface, which can be grasped by the color code, is very similar to the one obtained by exhaustive search in \figurename \ref{fig:exp_bump_gridsearch_top}. The efficacy of the proposed methodology can be appreciated by looking at the calibration suggested by Algorithm \ref{alg:bo}: they are sparse where the Gaussian Process estimated a worse performance index, and concentrated in a neighborhood of the optimum which lies in the same region identified by the exhaustive search. The full history of each experiment and associated performance index is shown in \figurename \ref{fig:exp_bump_bo_long_history}: the optimum calibration is found after only 8 experiments; one can notice how the first half of the experiments followed an exploitation strategy to reach the true minimum, whereas the following experiments were devoted to the exploration of the parameters space (consisting of higher values of the observed performance index).

\begin{figure}
    \centering
    \subfloat[]{\includegraphics[width=\columnwidth]{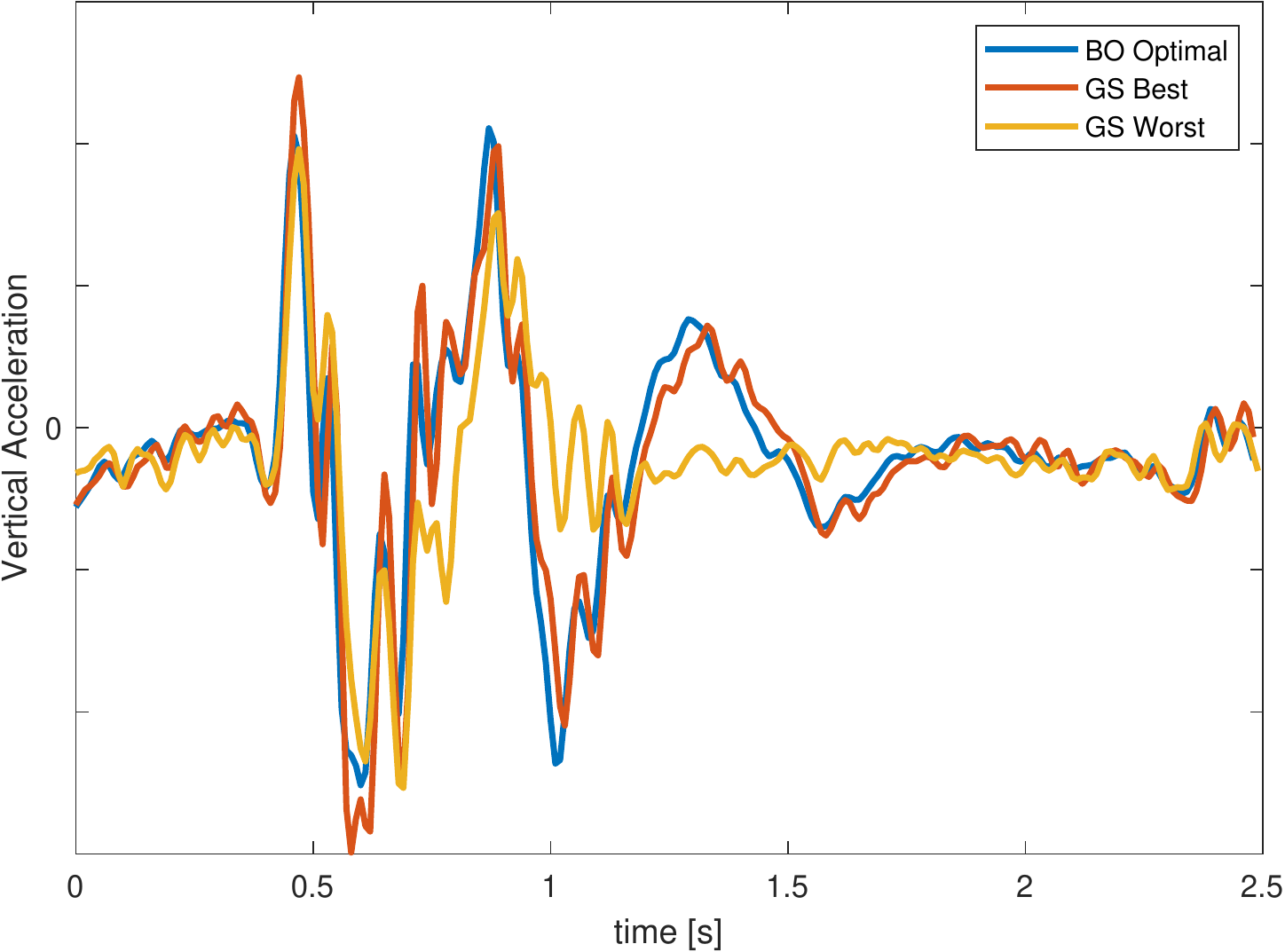} \label{fig:exp_bump_bo_long_comparison_az}}
    \hfil
    \subfloat[]{\includegraphics[width=\columnwidth]{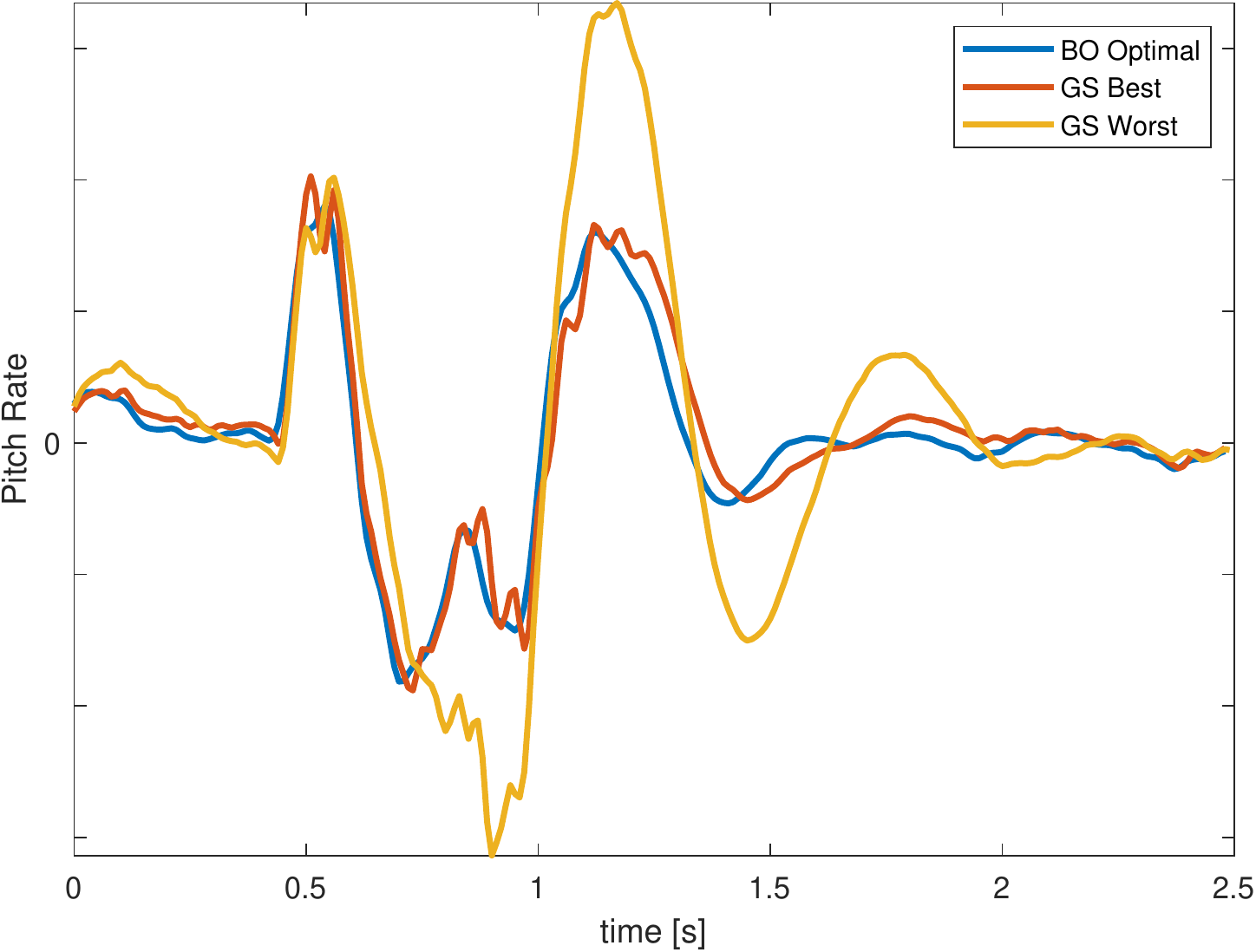} \label{fig:exp_bump_bo_long_comparison_pr}}
    \hfil
\caption{Measured vertical acceleration (a) and pitch rate (b) for \emph{Controller I} on the \emph{Speed Bump} with three calibrations.}
\label{fig:exp_bump_bo_long_comparison}
\end{figure}

\begin{figure}
	\centering
	\includegraphics[width=\columnwidth]{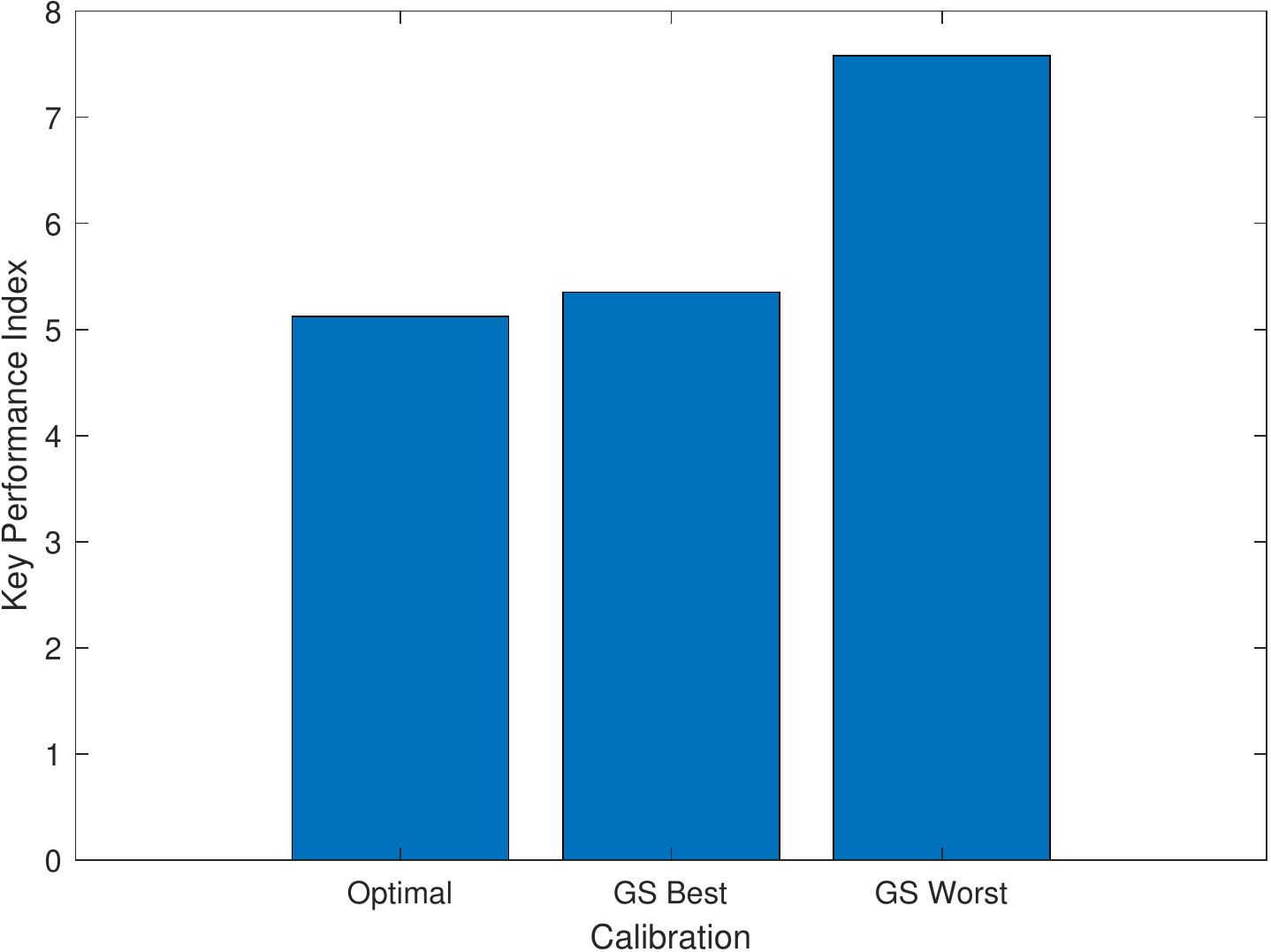}
	\caption{Performance Index for \emph{Controller I} on the \emph{Speed Bump} for three calibrations.}
	\label{fig:exp_bump_bo_long_comparison_kpi}
\end{figure}

The outcome of the calibration can be appreciated qualitatively in \figurename \ref{fig:exp_bump_bo_long_comparison} where the measured vertical acceleration and pitch rate are depicted for three configurations: the optimal calibration proposed by Algorithm \ref{alg:bo}, in blue, the optimal observed in the exhaustive search, in red, and the worst one, in yellow. As expected, the result is a trade-off between high vertical accelerations and oscillations due to the impact on the Speed Bump. A quantitative comparison of these three calibrations is shown in \figurename \ref{fig:exp_bump_bo_long_comparison_kpi} where it is possible to see that the optimal calibration outperforms the exhaustive search, since this latter is constrained on a coarse grid, whereas in the former case it is continuous in the parameters space.

It is therefore clear that the proposed search outperforms grid search with respect to both accuracy and number of experiments. For a qualitative appreciation: the grid search experiments took 3 hours of driving (and gas consumption), whereas our methodology only took 50 minutes.


\subsection{Test Road  for Controller I}

The test road in \figurename \ref{fig:exp_track} is characterized by a broad range of frequency excitation as discussed in Section \ref{sec:setup_exp}; hence, the optimal calibration must tackle the trade-off between a low-damping and high-damping configuration, compromising between low-frequency and high-frequency filtering of the suspension system. In this scenario, the exhaustive search could not be performed since it would have required a very long time.

\begin{figure}
	\centering
	\includegraphics[width=\columnwidth]{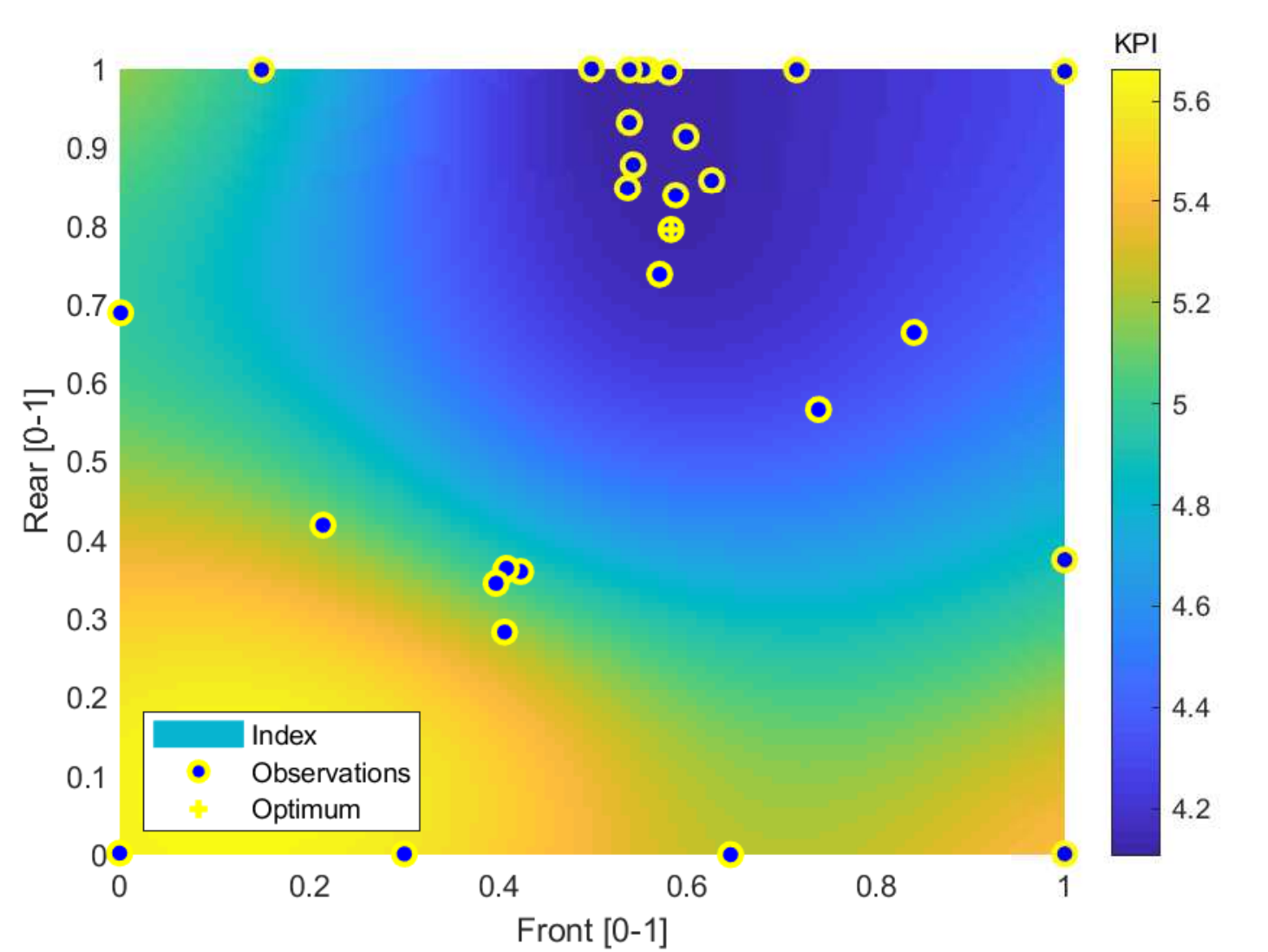}
	\caption{Estimation of the performance index for \emph{Controller I} on the \emph{Test Road} via a Gaussian Process.}
	\label{fig:exp_road_bo}
\end{figure}

\begin{figure}
	\centering
	\includegraphics[width=\columnwidth]{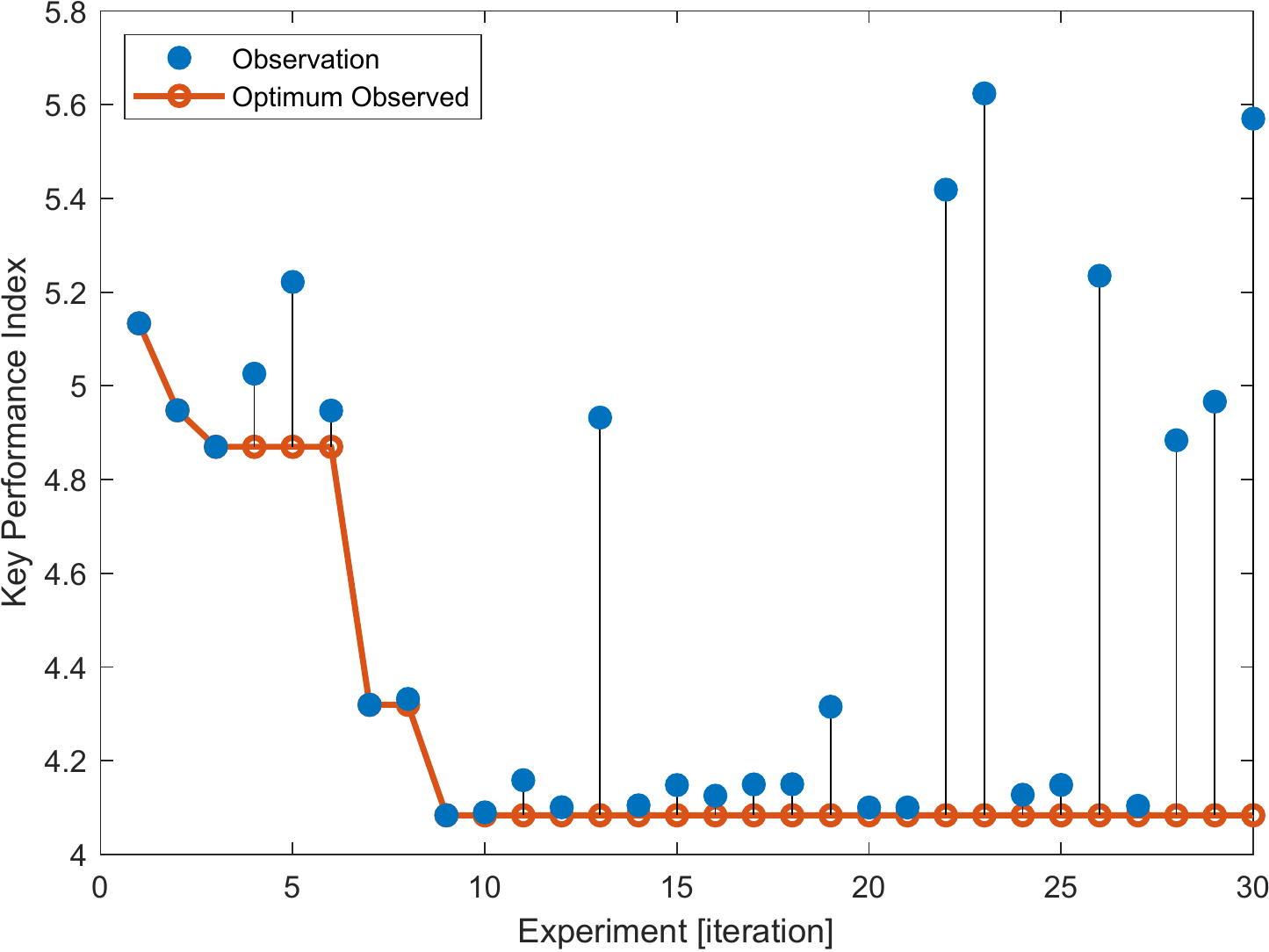}
	\caption{History of the performance index for each experiment for \emph{Controller I} on the \emph{Test Road}.}
	\label{fig:exp_road_bo_history}
\end{figure}

Following the experimental methodology presented in Section \ref{sec:methodology}: the test road is driven at the constant speed of $60 km/h$ along the same direction, the algorithm cuts and align the road segment of interest, computes the performance index, estimates the Gaussian Process from the observed data and eventually suggests a new calibration to be tested on the next iteration. The final result is depicted in \figurename \ref{fig:exp_road_bo} where the amplitude of the index follows the indicated color code, the proposed calibrations are represented by blue-yellow dots and the optimum calibration by a green cross. The solution is obviously different from the calibration obtained by driving over the speed bump since the road excitation is different. 
Similarly to the experiment on the speed bump, the optimal calibration is found after only 9 experiments, with the first half of iterations with a prevalence of exploitation over exploration, and the opposite for the second half, as shown in \figurename \ref{fig:exp_road_bo_history}.

\begin{figure}
    \centering
    \subfloat[]{\includegraphics[width=\columnwidth]{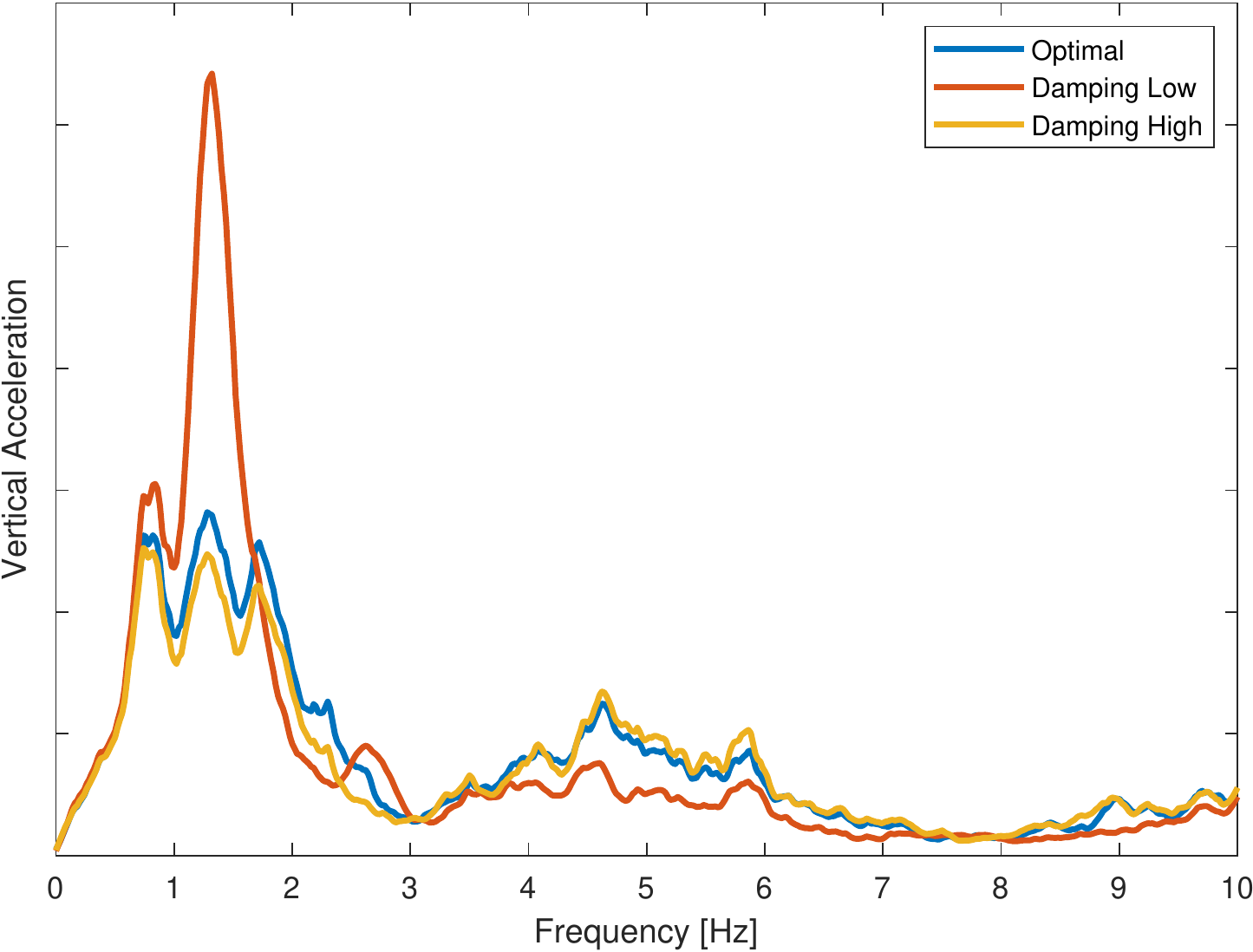} \label{fig:exp_road_bo_comparison_az}}
    \hfil
    \subfloat[]{\includegraphics[width=\columnwidth]{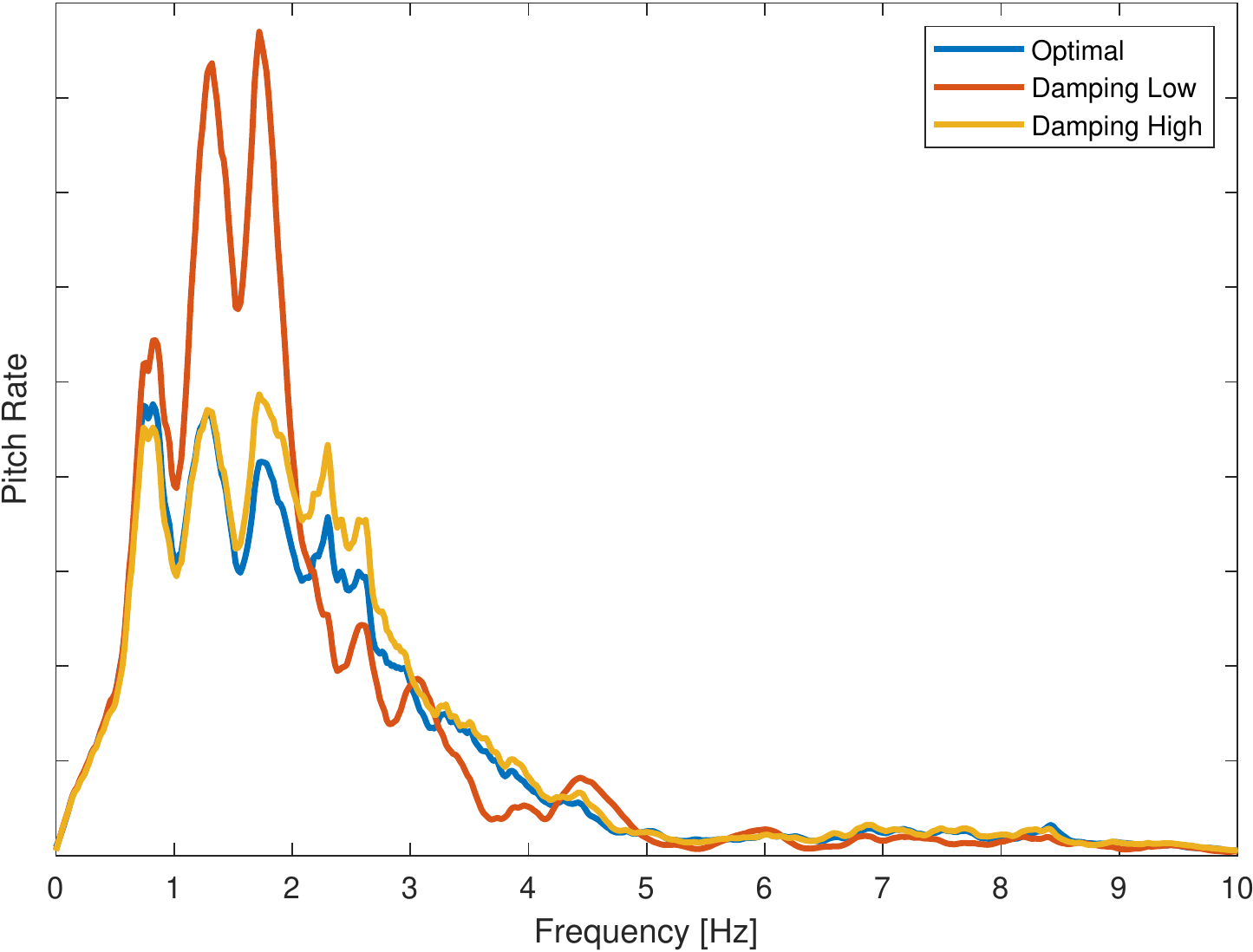} \label{fig:exp_road_bo_comparison_pr}}
    \hfil
\caption{Spectrum of the vertical acceleration (a) and pitch rate (b) for \emph{Controller I} on the \emph{Test Road} with three calibrations.}
\label{fig:exp_road_bo_comparison}
\end{figure}

The performance of the optimum calibration can be appreciated in terms of the Fourier transform of the vertical acceleration and pitch rate, as depicted in \figurename \ref{fig:exp_road_bo_comparison}. The result is compared against the passive configurations of minimum and maximum damping in order to better appreciate the trade-off: for lower frequencies the high damping has a superior damping effect, whereas at high-frequency vibrations are better attenuated by a low damping configuration; the optimum calibration is the best compromise between these two.

Now, the results on the calibration of the controller discussed in Section \ref{sec:controllers_semiactive} are presented. This controller is a real implementation of a production software consisting of 20 parameters; as discussed in Section \ref{sec:methodology}, the direct tuning on the vehicle of all parameters would not yield satisfactory results. 

Hence, a \emph{simulation phase} aimed at reducing to a subset of most significant parameters is needed; afterwards, this subset is calibrated on the vehicle during the \emph{experimental phase}.

\subsection{Simulation  for Controller II}

This \emph{phase} serves the purpose of identifying the most significant subset of parameters in order to reduce the full set consisting of 20 parameters. As discussed in Section \ref{sec:controllers}, the only knowledge on the system is the number of parameters and the range of values they can assume; no semantic knowledge is given \emph{a priori}.

Following the methodology discussed in Section \ref{sec:methodology}, 10 rounds of Bayesian Optimization are performed; each round is independent from the other since the initialization is performed at random and, subsequently, the exploration will be different. In this work, the optimization is performed using a Gaussian Process with Matérn Kernel and $N=100$ experiments; this number is chosen heuristically, since it has been observed that the optimization can always reach the minimum of the performance index within 100 experiments.

\begin{figure}
	\centering
	\includegraphics[width=\columnwidth]{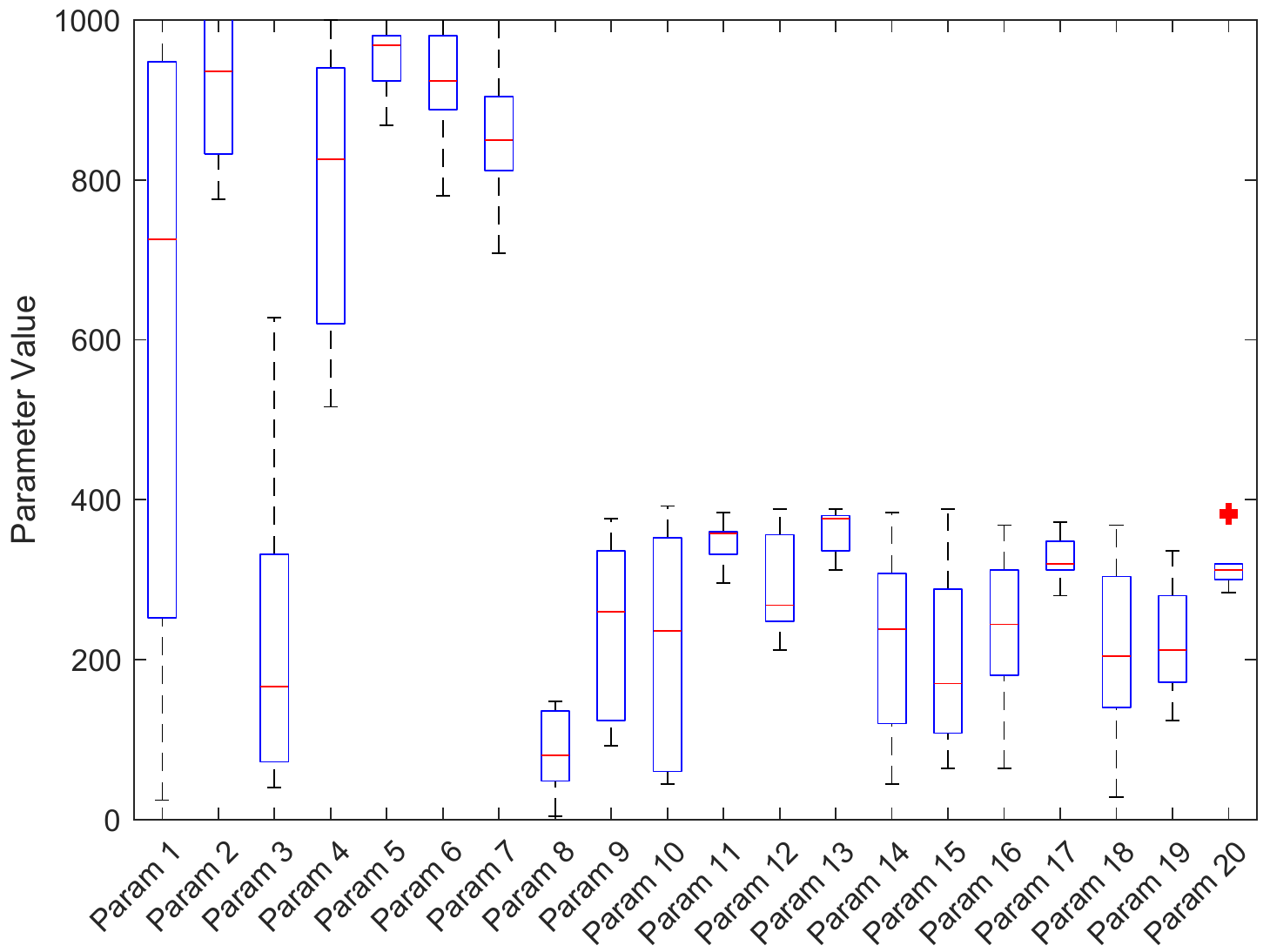}
	\caption{Distribution of the optimal value for each parameter of \emph{Controller II} in 10 repetitions of the optimization.}
	\label{fig:sim_boxplot}
\end{figure}

\begin{figure}
	\centering
	\includegraphics[width=\columnwidth]{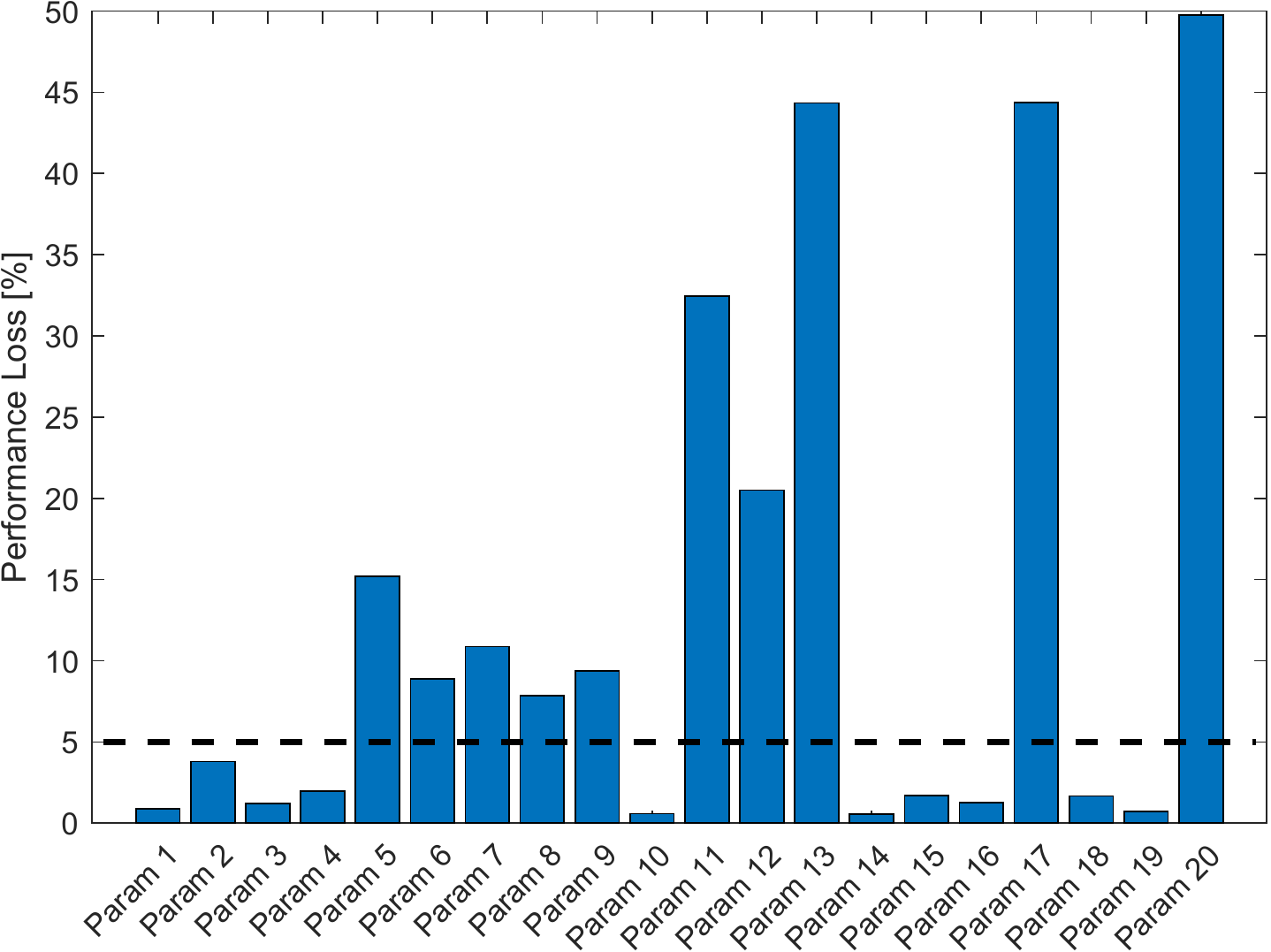}
	\caption{Maximum performance loss when each parameter is deviated from its optimal value.}
	\label{fig:sim_sensitivity}
\end{figure}

The result of these 10 optimizations, namely the optimal calibrations, is shown in \figurename \ref{fig:sim_boxplot}; one can observe that the optimal values for some parameters have a high variance, whereas others have a smaller one; we can draw three conclusions from this fact: 
\begin{enumerate}
    \item there exist more than one local minimum;
    \item these minima have a similar performance index;
    \item the high-variance parameters may have less effect on the performance index.
\end{enumerate}
Whilst the first two points are straightforward, in order to prove the latter we perform the sensitivity analysis proposed in Section \ref{sec:methodology}: starting from one of the 10 optimal calibrations, one parameter at the time is let vary between its extreme values; the maximum deviation from the minimum performance index is computed as in \eqref{eq:heuristics} and displayed in \figurename \ref{fig:sim_sensitivity}. This plot can be read as follows: if the value of \emph{Param 20} is mistakenly set, the performance of the software can suffer a 50\% detriment.

The most significant subset of parameters is chosen among those which exceed a $T=5\%$ loss in \figurename \ref{fig:sim_sensitivity}. Those parameters which have little effect on the performance index are set to a nominal value (the actual value is not important since the effect on the performance index is negligible).

\begin{figure}
    \centering
    \subfloat[]{\includegraphics[width=\columnwidth]{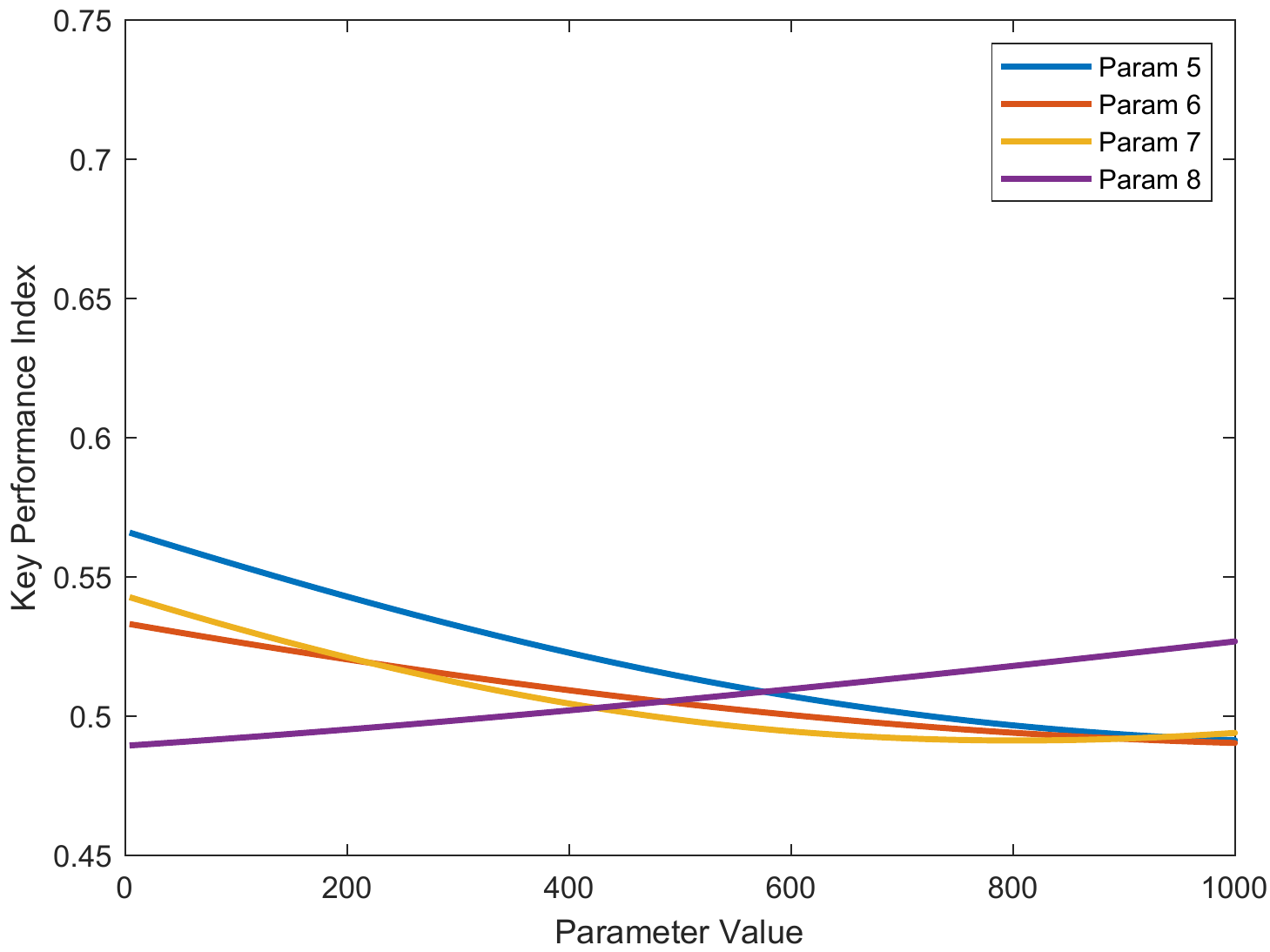} \label{fig:sim_sensitivity_1}}
    \hfil
    \subfloat[]{\includegraphics[width=\columnwidth]{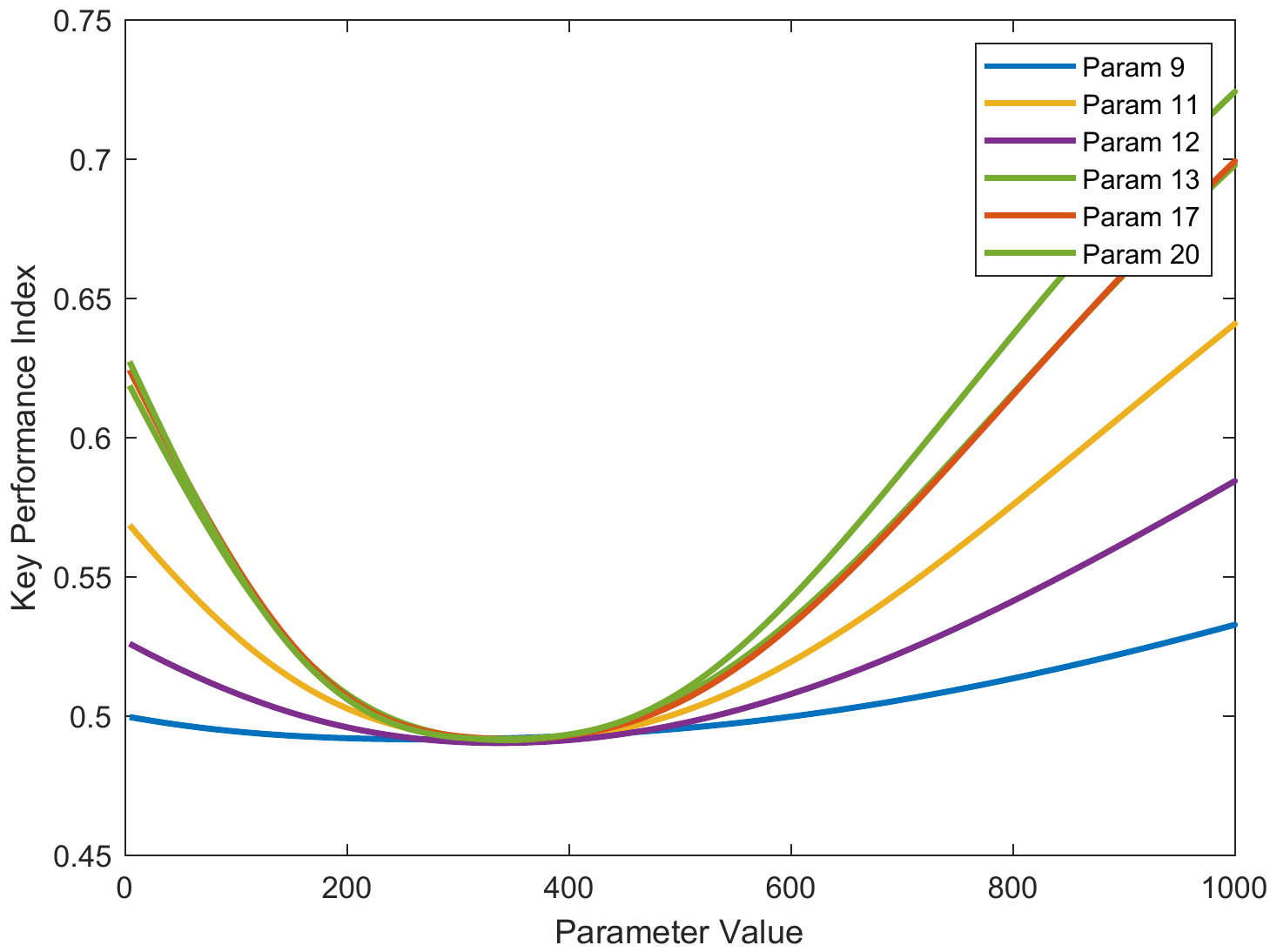} \label{fig:sim_sensitivity_2}}
    \hfil
\caption{Sensitivity of the performance index when one parameter at the time is let vary within its range.}
\label{fig:sim_sensitivity_1_2}
\end{figure}

To the aim of further reducing the number of parameters to be optimized directly on the vehicle, the aforementioned sensitivity analysis for the subset of most significant parameters is shown in \figurename \ref{fig:sim_sensitivity_1_2}, where each parameter is let vary in the range of definition. In \figurename \ref{fig:sim_sensitivity_1}, one can observe that the curves associated to \emph{Param 5}, \emph{Param 6} and \emph{Param 7} have the minimum located in a narrow region, whereas the optimum for \emph{Param 8} is found in the opposite direction with respect to these. Therefore, the parameters which show a similar optimal value are joined together, counting as a single \emph{meta-}parameter. Similarly, the parameters in \figurename \ref{fig:sim_sensitivity_2} can be considered as one.

Summarizing, only three \emph{meta-}parameters should be considered:
\begin{itemize}
    \item \emph{Param 5}, \emph{Param 6} and \emph{Param 7}
    \item \emph{Param 8}
    \item \emph{Param 9}, \emph{Param 11}, \emph{Param 12}, \emph{Param 13}, \emph{Param 17} and \emph{Param 20}.
\end{itemize}
All other parameters are set to a nominal value since they do not affect much the performance index.

Eventually, in order to demonstrate that this subset of \emph{meta-}parameters is indeed fully informative, we performed an additional optimization in this scenario. The results, compared in Table \ref{tab:optimum}, show that the performance loss of this constrained scenario is negligible.

\begin{table}[!t]
    \renewcommand{\arraystretch}{1.3}
    \caption{Performance index when considering parameter reduction}
    \label{tab:optimum}
    \centering
    \begin{tabular}{c c c}
    \hline
         Parameters & Optimal Value & Standard Deviation\\
         \hline \hline
         20 & 0.4917 & 0.0020 \\
         3  & 0.4926 & 0.0004 \\
         \hline
    \end{tabular}
\end{table}

\subsection{Experiments  for Controller II}
The output of the \emph{simulation phase} consists of a subset of only 3 \emph{meta-}parameters which have the most significant effect on the performance index.

These experiments were performed on the \emph{test road} in \figurename \ref{fig:exp_track} which is characterized by a broad range of frequency excitation as discussed in Section \ref{sec:setup}; the semi-active control logic is expected to perform better than the \emph{constant damping controller}, achieving superior performance for both low and high frequencies.  In  this  scenario,  the  exhaustive  search  could  not  be performed since it would have required a very long time and too many experiments in a tri-dimensional parameters space.

\begin{figure}
	\centering
	\includegraphics[width=\columnwidth]{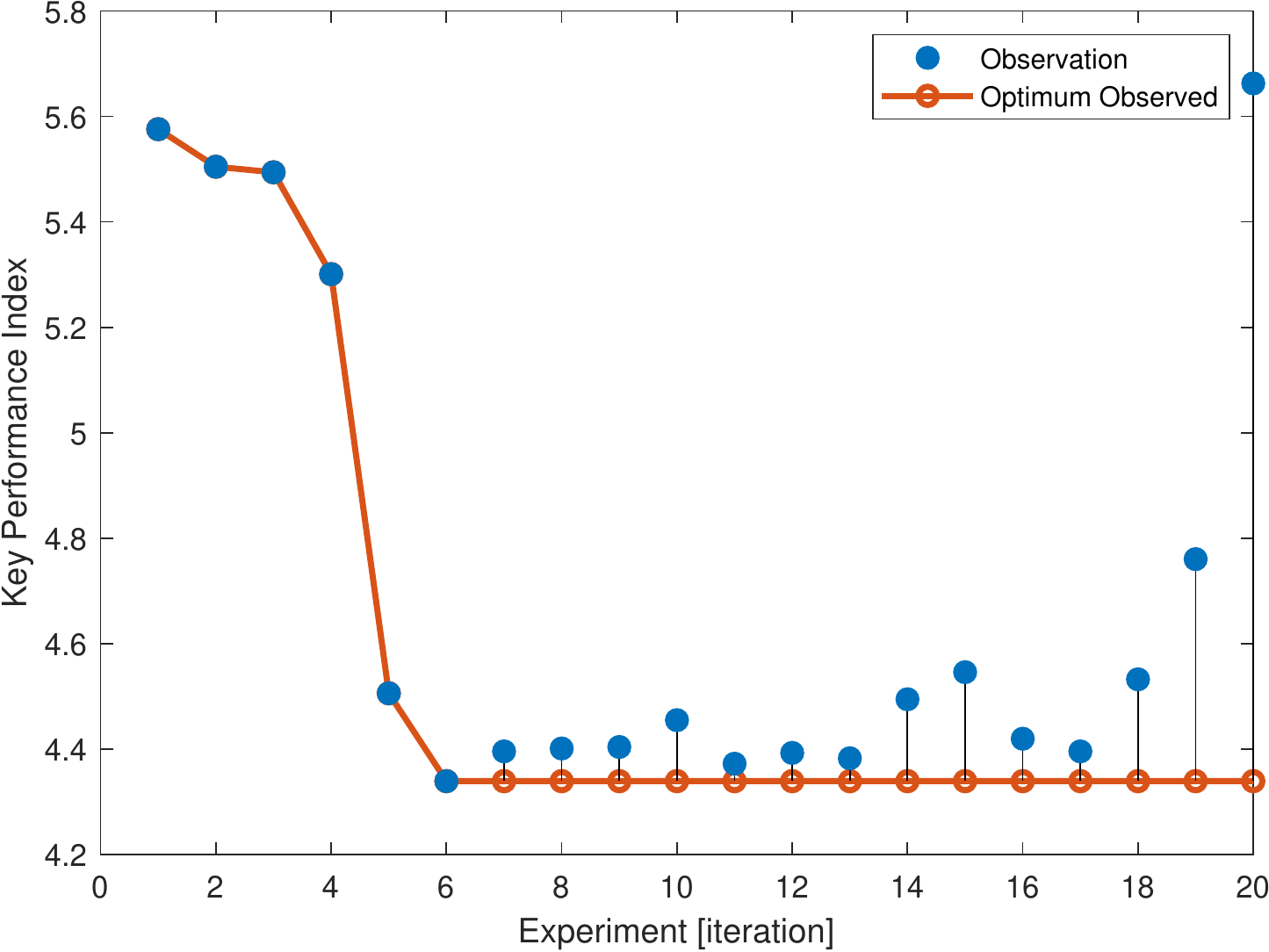}
	\caption{History of the performance index for each experiment for \emph{Controller II} on the \emph{Test Road}.}
	\label{fig:exp_road_bo_controller}
\end{figure}

Following the experimental methodology presented in Section \ref{sec:methodology}, the test road is driven at the constant speed of $60 km/h$ along the same direction, the algorithm cuts and align the road segment of interest, computes the performance index, estimates the Gaussian Process from the observed data and eventually suggests a new calibration to be tested on the next iteration. The optimal calibration is found after only 6 experiments as depicted in \figurename \ref{fig:exp_road_bo_controller} which is a remarkable results.

\begin{figure}
    \centering
    \subfloat[]{\includegraphics[width=\columnwidth]{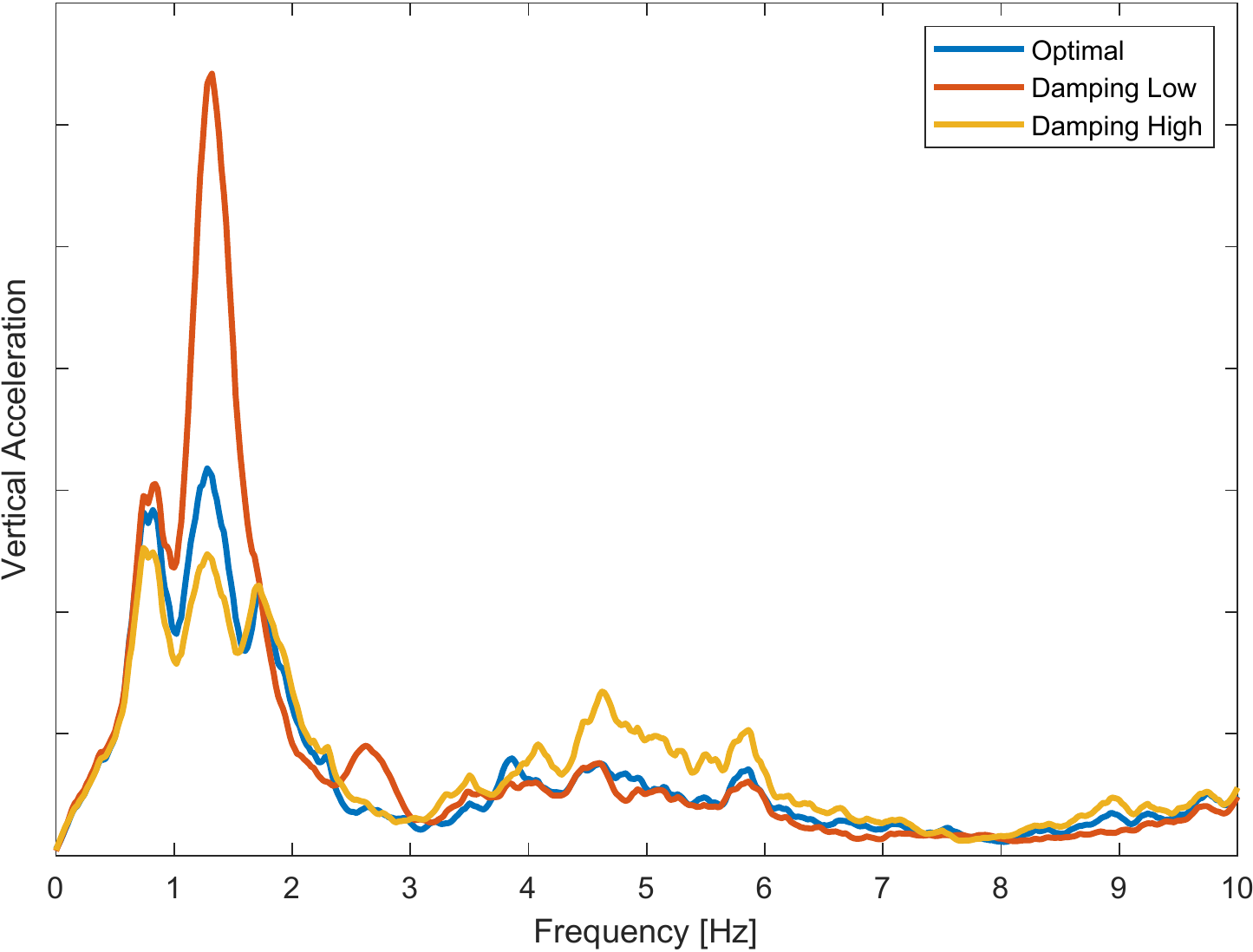} \label{fig:exp_road_bo_controller_comparison_az}}
    \hfil
    \subfloat[]{\includegraphics[width=\columnwidth]{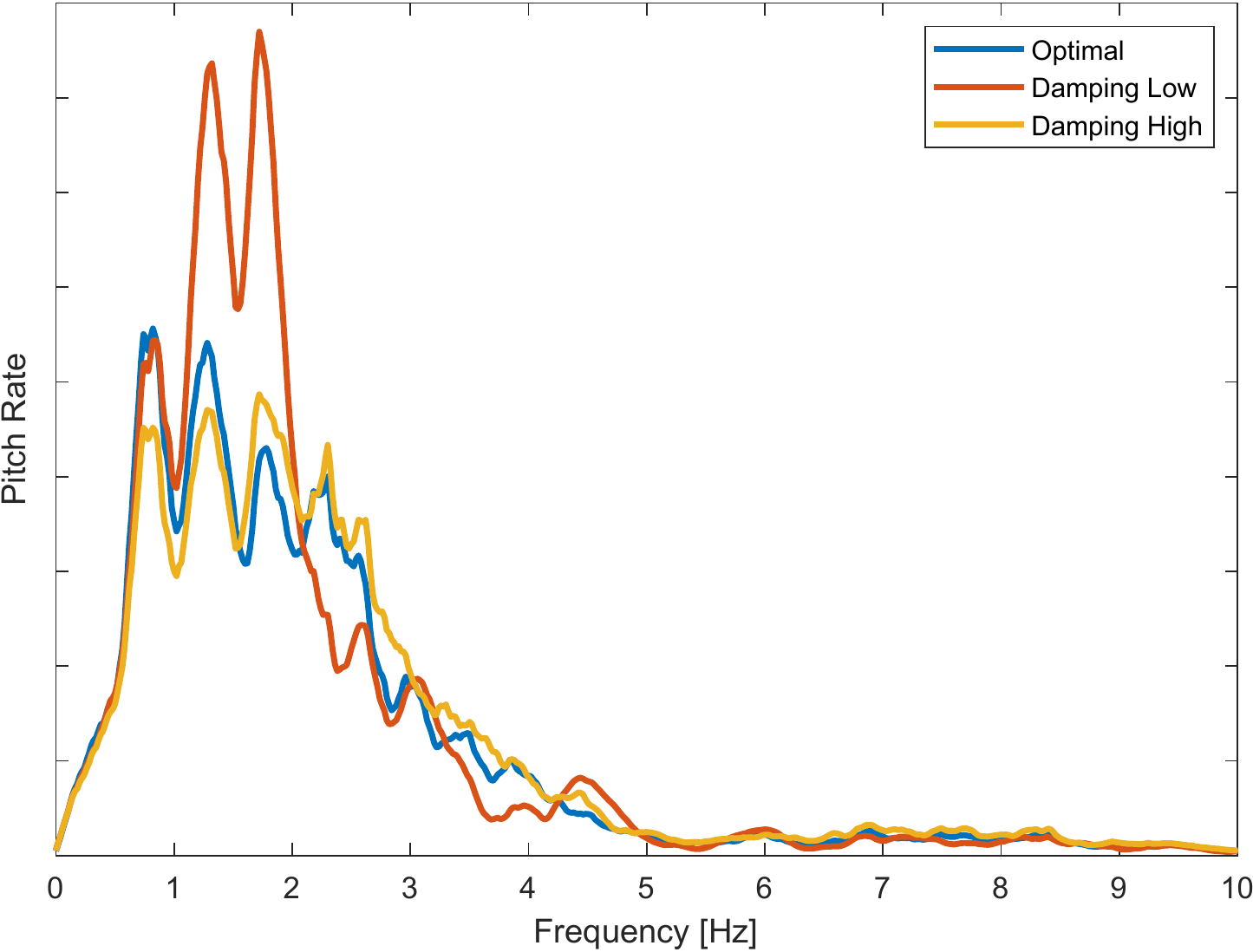} \label{fig:exp_road_bo_controller_comparison_pr}}
    \hfil
\caption{Spectrum of the vertical acceleration (a) and pitch rate (b) for \emph{Controller II} on the \emph{Test Road} with three calibrations.}
\label{fig:exp_road_bo_controller_comparison}
\end{figure}

The performance of the proposed calibration is shown in \figurename \ref{fig:exp_road_bo_controller_comparison} compared against the passive configurations in order to appreciate the goodness of the result: good damping of the heave resonance at $\sim 1 Hz$, and good filtering of the higher frequency vibrations at $\sim 5 Hz$ (mostly visible in \figurename \ref{fig:exp_road_bo_controller_comparison_az}).

The calibration experiments lasted only 1 hour which is a remarkably short duration with respect to the \emph{human} paradigm in \figurename \ref{fig:scheme_human}.

\section{Conclusions} \label{sec:conclusion}
In this work, we have shown how End-of-Line tuning of semi-active suspensions can be made fully automatic with BO tools and a properly designed experimental protocol. In particular, we have first formulated a suitable optimization problem inspired from the KPIs of interest, then we have designed an experimental procedure to feed the BO tool with informative data leading to the optimal calibration point. The distribution of the estimated parameters as well as the convergence of the optimization procedure have been also analyzed for the considered case studies.

Future works will be devoted to the use of reinforcement learning techniques to enforce a more efficient exploration of the parameter space and to enlarge the considered domain. Other maneuvers as well as suspensions technologies will be also accounted for. 


\section*{Acknowledgments}
The authors thank Hyundai Motor Company for providing the vehicle, without which this research would have not been possible.


\bibliographystyle{IEEEtran}
\bibliography{IEEEfull,bibliography}

%







\end{document}